\documentclass{aa}
\usepackage{graphicx}
\usepackage{amssymb}
\usepackage{natbib}
\bibpunct{(}{)}{;}{a}{}{,}

\newcommand{\Hii}{\ion{H}{ii} }
\newcommand{\Log}{\mbox{Log}}
\newcommand{\XMM}{XMM-{\em Newton} }
\newcommand{\chandra}{{\em Chandra} }

\newcommand{\lognlogs}{Log~$N$--Log~$S$ }
\newcommand{\lognlogsa}{Log~$N$--Log~$S$}
\begin{document}

\def\S{Sect.~}

\title{ The 2--10 keV luminosity as a Star Formation Rate indicator }

\author{P.\,Ranalli\inst{1} \and A.\,Comastri\inst{2} \and 
  G.\,Setti\inst{1} } 

\offprints{Piero Ranalli, \email{ranalli@bo.astro.it}}

\institute{
  Dipartimento di Astronomia, Universit\`a di Bologna,
  via Ranzani 1, I--40127 Bologna, Italy
\and 
  INAF -- Osservatorio Astronomico di Bologna,
  via Ranzani 1, I--40127 Bologna, Italy
}

\date{Received July 30, 2002; accepted November 4, 2002}

\abstract{ Radio and far infrared luminosities of star-forming
galaxies follow a tight linear relation. Making use of ASCA and
BeppoSAX observations of a well-defined sample of nearby star-forming
galaxies, we argue that tight linear relations hold between the X-ray,
radio and far infrared luminosities. The effect of intrinsic
absorption is investigated taking NGC\,3256 as a test case.  It is
suggested that the hard X-ray emission is directly related to the Star
Formation Rate.       Star formation processes may also account for most of
the 2-10 keV emission from LLAGNs of lower X-ray luminosities (for the
same FIR and radio luminosity).   Deep {\em Chandra} observations of a
sample of radio-selected star-forming galaxies in the Hubble Deep
Field North show that the same relation holds also at high
($0.2\lesssim z\lesssim 1.3$) redshift. The X-ray/radio relations also
allow a derivation of X-ray number counts up to very faint fluxes from
the radio \lognlogsa, which is consistent with current limits and
models. Thus the contribution of star-forming galaxies to the X-ray
background can be estimated.


}

\maketitle

\section{Introduction}

Radio continuum and far infrared (FIR) luminosities of star-forming
galaxies are known to show a {\em tight linear} relationship spanning
four orders of magnitude in luminosity
\citep{pranalli-E3:vdK73,pranalli-E3:dejo85,pranalli-E3:cond92} and up
to a redshift $\sim 1.3$ \citep{garrett02}.  This is interpreted as
due to the presence of massive, young stars embedded in dust: a
fraction of their UV radiation is absorbed by dust grains and
reradiated in the infrared band, while supernova explosions may
accelerate the electrons producing at radio wavelengths the observed
synchrotron emission \citep{pranalli-E3:hp75,pranalli-E3:helou}.
Since massive ($M\gtrsim 5 ~\rm M\sun$) stars are short-lived, these
luminosities are assumed to be indicators of the global Star Formation
Rate (SFR) in a galaxy.  Following \citet{pranalli-E3:cond92} and
\citet{pranalli-E3:kenn98}, the relation between SFR (referred to
stars with $M>5$ M$\sun$) and radio/FIR luminosities can be written
as:
\begin{eqnarray}
{\rm SFR}=\frac{L_{\rm 1.4 GHz}}{4.0\cdot 10^{28}}~ \mbox{M$\sun$/yr}
	 \label{pranalli-E3:eqsfr1}  \\
{\rm SFR}=\frac{L_{\rm FIR}}{2.2\cdot 10^{43}}~  \mbox{M$\sun$/yr}
	\label{pranalli-E3:eqsfr2}
\end{eqnarray}
with the FIR flux defined after \citet{pranalli-E3:helou} as:
\begin{equation}
{\rm FIR}=1.26\cdot 10^{-11} (2.58 S_{60\mu} + S_{100\mu})~ \mbox{erg s$^{-1}$ cm$^{-2}$}
\end{equation}
where $L_{\rm 1.4 GHz}$ is in erg\,s$^{-1}$\,Hz$^{-1}$, $L_{\rm FIR}$
in erg s$^{-1}$ and infrared fluxes are in Jy.

Star-forming galaxies are also luminous sources of X-ray emission, due
to a number of High Mass X-ray Binaries (HMXB), young supernova
remnants, and hot plasmas associated to star-forming regions and
galactic winds \citep{fabbiano89}.  A non linear ($L_{\rm X} \propto
L_{\rm FIR}^{0.6}$) and much scattered (dispersion of about 2 dex)
relation was found between FIR and soft (0.5--3.0 keV) X-ray
luminosities of IRAS-bright and/or interacting/peculiar galaxies
measured by the {\em Einstein} satellite \citep{gp90}. A somewhat
different result was found by \citet{djf92}, i.e. a linear relation
between FIR and 0.5--4.5 keV luminosities for a sample of starburst
galaxies observed by {\em Einstein}. A large number of upper limits to
the X-ray flux (12 upper limits vs.~11 detections for \citealt{gp90})
along with high uncertainties in the X-ray and FIR fluxes may explain
this discrepancy. Moreover, these studies suffered by the lack of
knowledge about spectral shapes and internal absorption in star
forming galaxies caused by the limited sensitivity and spectral
capabilities of the IPC detector onboard {\em Einstein}.

%PAPER OUTLINE
Here, with the high sensitivity and the broad-band spectral
capabilities of the ASCA and BeppoSAX satellites, we extend these
studies to the 2--10 keV band which is essentially free from
absorption. In the following paragraphs a sample of nearby star
forming galaxies is assembled (\S\ref{campione_section}) and linear
relations among radio, FIR and both soft and hard X-ray luminosities
are found (\S\ref{correlazioni_section}).  Possible biases are
discussed and the use of X-ray luminosities as a SFR indicator is
proposed (\S\ref{biases_section}). In \S\ref{hdf_section} we present a
study of star-forming galaxies in the Hubble Deep Field North and test
the validity of the X-ray SFR law. Implications for the contribution
of star-forming galaxies to the X-ray counts and background are
discussed in \S\ref{fondox_section}.

Throughout this paper, we assume H$_0=50$ and q$_0=0.1$.

\section{The local sample}       \label{campione_section}

The atlas of optical nuclear spectra by \citet{pranalli-E3:hfs97}
(hereafter HFS97) represents a complete spectroscopic survey of
galaxies in the Revised Shapley-Ames Catalog of Bright Galaxies
(RSA; \citealt{pranalli-E3:RSA}) and in the Second Reference Catalogue
of bright galaxies (RC2; \citealt{rc2}) with declination $\delta >
0\degr$ and magnitude $B_T<12.5$.  Optical spectra are classified in
HFS97 on the basis of line intensity ratios according to
\citet{pranalli-E3:vo87}; galaxies with nuclear line ratios typical of
star-forming systems are labeled as ``\Hii nuclei''.
This sample of \Hii galaxies contains only spirals and irregulars from Sa
to later types, except for a few S0 which were excluded from our
analysis since their properties resemble more those of elliptical
galaxies.

A cross-correlation of the HFS97 sample with the ASCA
archive gives 18 galaxies clearly detected in the 2--10
keV band with the GIS instruments.  Four additional objects in
the field of view of ASCA observations were not detected: the 2--10
keV flux upper limits are too loose to add any significant
information, and thus we did not include them in the sample.
The cross-correlation of the HFS97 sample with the BeppoSAX archive
does not increase the number of detections. When a galaxy was observed
by both satellites, we chose the observation with better quality data.

Far infrared fluxes at 60$\mu$ and 100$\mu$ were taken from the IRAS
Revised Bright Galaxy Sample (RBGS, Mazzarella et al., in
preparation) which is a reprocessing of the final IRAS archive with
latest calibrations.  While the RBGS measurements should be more
accurate, we checked that the use of the older catalogue of IRAS
observations of large optical galaxies\footnote{Blue-light isophotal
major diameter ($D_{25}$) greater than $8^\prime$.}  by
\citet{rice88}, coupled with the Faint Source Catalogue (FSC,
\citealt{fsc}) for smaller galaxies, does not significantly change our
statistical analysis. FIR fluxes for NGC\,4449 were taken from
\citet{rush}.  Radio (1.4\,GHz) fluxes were
obtained from the \citet{pranalli-E3:cond90,pranalli-E3:cond96}
catalogues (except for NGC\,4449, taken from \citealt{haynes75}).
Distances were taken from \citet{tully} and corrected for the adopted
cosmology.

% flusso limite iras:
%  The FSC contains data for 173,044 point sources
%  in unconfused regions with flux densities typically above 0.2 Jy at
%  12, 25, and 60 microns, and above 1.0 Jy at 100 microns.

% ngc4449 non c'e' nel FSC (credo sia nel faint source reject ma e' un
% casino trovarlo) perche' per lo FSC servivano almeno 6 scan mentre lei
% ne ha solo 2
%
% cmq c'e' nel 12 micron sample di rush et al., ed il loro flusso infrarosso
% concorda con quello di mazzarella joe

Part of the X-ray data have already been published; in the cases where
published data were not available in a form suitable for our analysis,
the original data were retrieved from the archive and reduced following
standard procedures and with the latest available calibrations.
Images and spectra were extracted from the pipeline-screened event
files.  The images were checked against optical (Digital Sky Survey)
and, where available, radio (1.4 GHz) images in order to look for
possible source confusions. Fluxes were calculated in the 0.5--2.0 and
2--10 keV bands from best-fit spectra for the GIS2 and GIS3
instruments and corrected for Galactic absorption only. The uncertainty
on the fluxes is of the order of 10$\%$. Depending on
the quality of data, the best-fit spectrum is usually represented by a
two-component model with a thermal plasma plus a power-law or just a
power-law.

The galaxy IC\,342 has shown some variability, mainly due to a burst
in 1993 of an ultraluminous X-ray binary.      M82 is also a variable
source.   For each galaxy we summarize
in the Appendix the results from several X-ray observations and
estimate a time-averaged luminosity.

 One object
(\object{M33}) was not included in the sample since its broad-band
(0.5--10 keV) X-ray nuclear spectrum is dominated by a strong variable
source (\object{M33 X-8}) identified as a black hole candidate
\citep{pranalli-E3:parmar01}. % Although M33 is identified by HFS97 as
%an \Hii nucleus, it has a very low SFR ($\sim 0.009$ M$\sun$/yr) so
%that the spectral signatures related to star formation can be easily
%hidden by a single powerful source.

% completezza
Therefore, our sample (hereafter local sample) consists of the 17
galaxies listed in Table \ref{campione}.  Since it is not complete in
a strict sense due to the X-ray selection, we have checked for its
representativeness with reference to the SFR. The median SFR values
for HFS97 is SFR$_{\rm med} \sim$ 1.65 M$\sun$/yr.  Considering
objects with SFR $> {\rm SFR}_{\rm med}$ there are 14 galaxies in the
local sample out of 98 in HFS97 (14$\%$), while there are 3 objects
with SFR $<{\rm SFR}_{\rm med}$ (3$\%$). Thus the high luminosity tail
is better sampled than the low luminosity one.

%cross-correlated the HFS97 sample with the \citet{rice88} and the FSC
%catalogues, obtaining a complete homogeneous sample of 196 nearby
%($z<0.01$) star-forming galaxies with known SFR (notice that the FSC
%only covers the sky with galactic latitude $|b|>10\degr$ and is
%complete down to limiting fluxes of 0.2\,Jy at $60\mu$ and 1.0\,Jy at
%$100\mu$).
%
%In our X-ray-detected sample 14 galaxies out of 120 in the complete
%sample ($12\%$) have a ${\rm SFR}>1~ {\rm M}\sun/$yr, and 11 out of 72
%have a ${\rm SFR}>3~ {\rm M}\sun/$yr ($15\%$). 

%All
%statistical tests presented here are performed on this sample unless
%otherwise stated.

We also include data for 6 other well-known starburst
galaxies which were not in the HFS97 survey because they are in the
southern emisphere. On the basis of their line intensity ratios%
\footnote{References: NGC\,55 - \citet{webster83}; NGC\,253, 1672 \&
1808 - \citet{kewley}; NGC\,3256 - \citet{moran99}; Antennae -
\citet{rubin70}, \citet{crimea}.  } they should be classified as \Hii
nuclei.  In Table \ref{campione} we label them as supplementary
sample.

% ~~~~~~~~~~~~~~~~~~~~~~~~~~~~~
%  TABELLA: DATI LOCAL SAMPLE
% ~~~~~~~~~~~~~~~~~~~~~~~~~~~~~
\begin{table*}[p]	
\centering\begin{tabular}{lcccccccccc}
\multicolumn{11}{c}{\sc Fluxes and Luminosities: Main Sample} \\
\\
%  & &\multicolumn{2}{c}{\sc 0.5--2.0 keV} &\multicolumn{2}{c}{\sc 2.0--10 keV} 
%&\multicolumn{2}{c}{\sc FIR} &\multicolumn{2}{c}{\sc 1.4 GHz} \\
{\sc Galaxy} &{D} &$F_{0.5-2}$ &$L_{0.5-2}$ &$F_{2-10}$ &$L_{2-10}$ &$F_{\rm FIR}$ &$L_{\rm FIR}$ &$F_{1.4}$ &$L_{1.4}$  &{\sc Refs.} \\
\hline

\object{M82}*    &7.8  &97   &7.1   &290  &21    &67   &49   &7.7  &5.6   &1      \\
\object{M101}    &8.1  &5.4  &0.42  &6.8  &0.54  &6.0  &4.7  &0.75 &0.59  &this work \\
\object{M108}    &21   &4.4  &2.3   &6.0  &3.2   &2.0  &11   &0.31 &1.63  &this work       \\
\object{NGC891}  &14   &8.3  &2.1   &19   &4.8   &4.5  &11   &0.70 &1.74  &this work\smallskip\\ 
\object{NGC1569} &2.4  &5.4  &0.037 &2.2  &0.015 &2.5  &0.17 &0.41 &0.028 &2      \\
\object{NGC2146} &26   &8.2  &6.5   &11   &9.0   &7.3  &59   &1.1  &8.7   &3      \\
\object{NGC2276} &55   &2.1  &7.6   &4.4  &16    &0.85 &31   &0.28 &10    &this work       \\
\object{NGC2403} &6.3  &16   &0.77  &9.3  &0.44  &2.7  &1.3  &0.33 &0.16  &this work\smallskip\\
\object{NGC2903} &9.4  &7.9  &0.84  &7.0  &0.74  &3.7  &4.0  &0.41 &0.43  &this work       \\
\object{NGC3310} &28   &7.4  &6.9   &2.1  &2.0   &1.7  &16   &0.38 &3.6   &4      \\
\object{NGC3367} &64   &1.8  &8.7   &1.6  &7.9   &0.38 &19   &0.10 &4.9   &this work       \\
\object{NGC3690} &69   &5.7  &33    &11   &62    &5.3  &310  &0.66 &38    &4\smallskip \\
\object{NGC4449} &4.5  &8.3  &0.20  &4.8  &0.12  &1.9  &0.47 &0.6  &0.1   &5      \\
\object{NGC4631} &10   &9.4  &1.2   &9.3  &1.2   &4.9  &6.2  &1.2  &1.5   &6      \\
\object{NGC4654} &25   &0.6  &0.4   &0.9  &0.66  &0.93 &7.1  &0.12 &0.89  &this work       \\
\object{NGC6946} &8.2  &30   &2.5   &12   &0.97  &7.9  &6.4  &1.4  &1.1   &this work       \\
\object{IC342}   &5.9  &18   &0.73  &110  &4.4   &11   &4.4  &2.3  &0.92  &this work   \\
\hline
\\ \\
\multicolumn{11}{c}{\sc Supplementary Sample} \\
\\
%{\sc Galaxy} &{\sc Dist} &F052 &L052 &F210 &L210 &FIR &LFIR &S14 &L14 &{\sc Refs.} \\
\hline
\object{NGC55}    &1.9  &18   &0.082 &6.8  &0.031&4.7  &0.21 &0.38 &0.017 &6      \\
\object{NGC253}*   &4.5  &25   &0.59  &50   &1.2  &49   &12   &5.6  &1.3  &1      \\
\object{NGC1672}  &22   &5.8  &3.2   &6.1  &3.4  &2.3  &13   &0.45 &2.5   &this work       \\
\object{NGC1808}  &16   &6.5  &2.0   &7.6  &2.4  &5.3  &17   &0.52 &1.6   &this work       \\
\object{NGC3256}  &56   &9.0  &34    &6.2  &23   &4.8  &180  &0.66 &25    &7      \\
\object{Antennae} &38   &7.2  &13    &5.3  &9.2  &2.6  &45   &0.57 &9.9   &8      \\
\hline

\end{tabular}
%\multicolumn{3}{c}{\sc Local Sample}\\ 
%\hline\hline
%
%\object{M82}*     &\object{NGC\,2276} &\object{NGC\,4449}\\
%\object{M101}    &\object{NGC\,2403} &\object{NGC\,4631}\\
%\object{M108}    &\object{NGC\,2903} &\object{NGC\,4654}\\ 
%\object{NGG\,891}  &\object{NGC\,3310}&\object{NGC\,6946}\\
%\object{NGC\,1569} &\object{NGC\,3367} &\object{IC\,342}  \\
%\object{NGC\,2146} &\object{NGC\,3690} \\
%\\
%\multicolumn{3}{c}{\sc Supplementary Sample }\\
%\hline\hline
%\object{NGC\,55}	&\object{NGC\,1672} &\object{NGC\,3256}\\
%\object{NGC\,253}*	&\object{NGC\,1808} &\object{Antennae}\\
%\end{tabular}
\caption{Data for galaxies in our local samples. All galaxies were observed with
ASCA, except those marked with * observed by BeppoSAX.
Distances in Mpc;
X-ray fluxes in $10^{-13}$ erg s$^{-1}$ cm$^{-2}$, FIR fluxes in 
$10^{-9}$ erg s$^{-1}$ cm$^{-2}$ and radio fluxes in Jy; X-ray luminosities in
$10^{40}$ erg s$^{-1}$, FIR luminosities in $10^{43}$ erg s$^{-1}$
and radio luminosities in $10^{29}$ erg s$^{-1}$ Hz$^{-1}$. The uncertainty
on fluxes and luminosities is of the order of 10$\%$. \newline
References: 1 \citet{cappi99}; 2 \citet{dellaceca96};
3 \citet{dellaceca99}; 4 \citet{zezas98}; 5 \citet{dellaceca97};
6 \citet{dahlem98}; 7 \citet{moran99}; 8 \citet{sansom96}. \newline
\label{campione}
}
\end{table*}

\section{The radio/FIR/X-rays correlation}         \label{correlazioni_section}

\begin{figure*}[!t]    % firsx e radiosx
  \begin{center} 
      \includegraphics[width=0.49\textwidth]{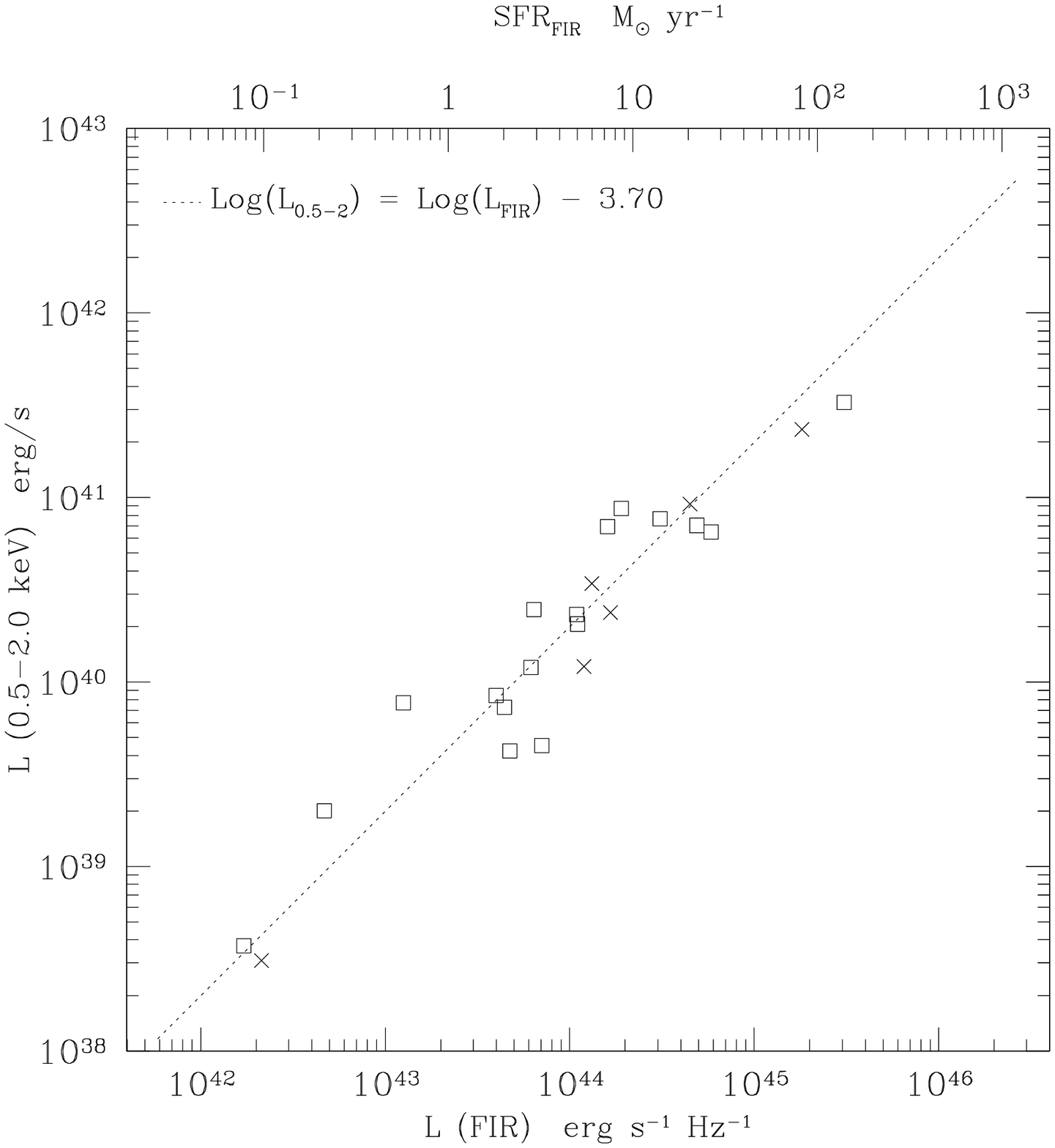}
      %\hskip.8cm
      \includegraphics[width=0.49\textwidth]{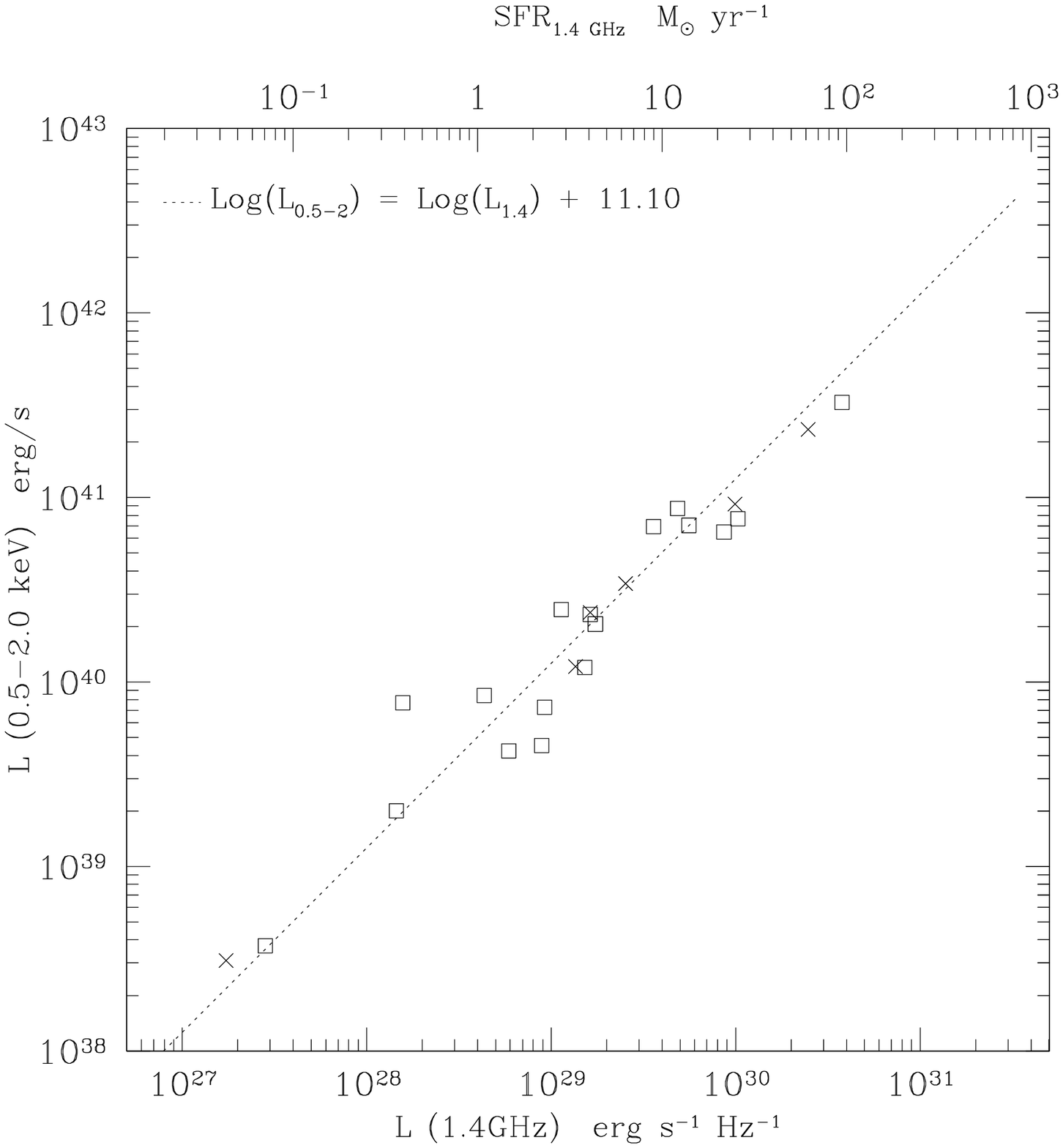}
  \end{center}
  \caption {The 0.5--2.0 keV luminosity of local star-forming galaxies
  vs.~radio and FIR ones. 
  Squares: local sample; crosses: supplementary sample;
    dotted lines: eqs.(\ref{eq:firx}, \ref{eq:radiox}).
  \label{pranalli-E3_fig:fig1}
  }
\end{figure*}

As a preliminary test, we perform a least-squares analysis for the
well-known radio/FIR correlation, which yields
\begin{equation}\label{pranalli-E3:radiofir}
\Log (L_{\rm FIR}) = (0.98\pm 0.06) ~\Log (L_{\rm 1.4}) 
     +15.4\pm 1.6
\end{equation}

The dispersion around the best-fit relation is given as the
estimate $s$ of the standard deviation $\sigma$:
\begin{equation}
s = \frac{1}{N-\nu} \cdot \sqrt {\sum \left( \Log~ L_{\rm obs}-\Log~ L_{\rm pred}
\right)^2 }     \label{eqvarianza}
\end{equation}
where $\nu$ is the number of free parameters
and $N$ is the number of points in the fit),
$L_{\rm pred}$ is the luminosity expected from the best fit
relation and $L_{\rm obs}$ the observed one.
For the radio/FIR correlation
(Eq.~\ref{pranalli-E3:radiofir})  we find $s=0.18$.

Following \citet{pranalli-E3:helou} we also calculate the mean ratio
$q$ between the logarithms of FIR and radio fluxes, obtaining
$q\simeq 2.2\pm 0.2$. This value is consistent
with the mean $q=2.34\pm 0.01$ for the 1809 galaxies in
the IRAS 2\,Jy sample by \citet{pranalli-E3:yrc}.

%\begin{equation}
%q=\Log ({\rm FIR}/3.75\cdot 10^{12}~{\rm Hz}/S_{1.4})
%\end{equation}
%where $3.75\cdot 10^{12}~{\rm Hz}$ is the frequency at $80\mu$, and
%$S_{1.4}$ is in erg/s/cm$^2$.

\subsection*{Soft X-rays}

A  test for the soft X-ray/FIR/radio relations
(Fig.~\ref{pranalli-E3_fig:fig1}) yields \\
\begin{eqnarray}
\Log (L_{0.5-2}) &= &(0.87\pm 0.08) ~\Log (L_{\rm FIR}) 
	+2.0\pm 3.7 \label{eq:firx1} \\
\Log (L_{0.5-2}) &= &(0.88\pm 0.08) ~\Log (L_{\rm 1.4})
	 +14.6\pm 2.2 \label{eq:radiox1} \medskip
\end{eqnarray}\\
with $s\simeq 0.26$ and 0.24 respectively.

Our result is consistent with the $L_{0.5-4.5}\propto L_{\rm
FIR}^{0.95\pm 0.06}$ relation found by \citet{djf92} for normal and
starburst galaxies from the IRAS Bright Galaxy Sample, but it is only
marginally consistent with the much flatter and more dispersed
relationship obtained by \citet{gp90} for a sample of IRAS selected
galaxies ($L_{0.5-3 \rm keV} \propto L_{60\mu}^{0.62\pm 0.14}$) and
for a sample of starburst/interacting galaxies ($L_{0.5-3 \rm keV}
\propto L_{60\mu}^{0.70\pm 0.12}$).

The inclusion of the objects of the supplementary sample (Table~%
\ref{campione}) does not significantly change
the slopes, i.e.  $L_{0.5-2.0 \rm keV}\propto L_{\rm
FIR}^{0.88\pm 0.07}$; likewise, if we use the $60\mu$ luminosity
instead of FIR, we obtain $L_{0.5-2.0 \rm keV}\propto
L_{60\mu}^{0.85\pm 0.07}$. 

By assuming an exactly linear slope, the best fit relations 
for the local (local+supplementary) sample become:\\
\begin{eqnarray}
\Log (L_{\rm 0.5-2}) &= &\Log (L_{\rm FIR}) - 3.68~~ (3.70) \label{eq:firx}\\
\Log (L_{\rm 0.5-2}) &= &\Log (L_{\rm 1.4}) + 11.08~~ (11.10) 
   \label{eq:radiox}
\end{eqnarray}
with $s \simeq 0.27$ and 0.24 respectively.

%ora l'F-test
%in grassetto perche' e' post-referee
%{\bf The F-test for an additional term in best-fits can be used to 
%compare the significance of the free-slope and fixed-slope fits.
%Under the assumption of equal uncertainties for each point in the fit,
%it can be shown that
%the F parameter reduces to 
%\begin{equation}
%F=\frac{s_1^2\frac{\nu+1}{\nu} - s_2^2}{s_2^2} \cdot \nu
%\end{equation}
%where $s_1^2$ and $s_2^2$ are the variance estimates for the free-slope
%and fixed-slope fits (eq.~\ref{eqvarianza}), and
%$\nu$ is the degrees of freedom for free-slope fit.
     By applying an F-test we find that the free-slope fits are not
significantly better than those with the linear slope, the improvement
being significant only at the $1\sigma$ level.

\begin{figure*}[!t]    % firx e radiox
  \begin{center} 
      \includegraphics[width=0.49\textwidth]{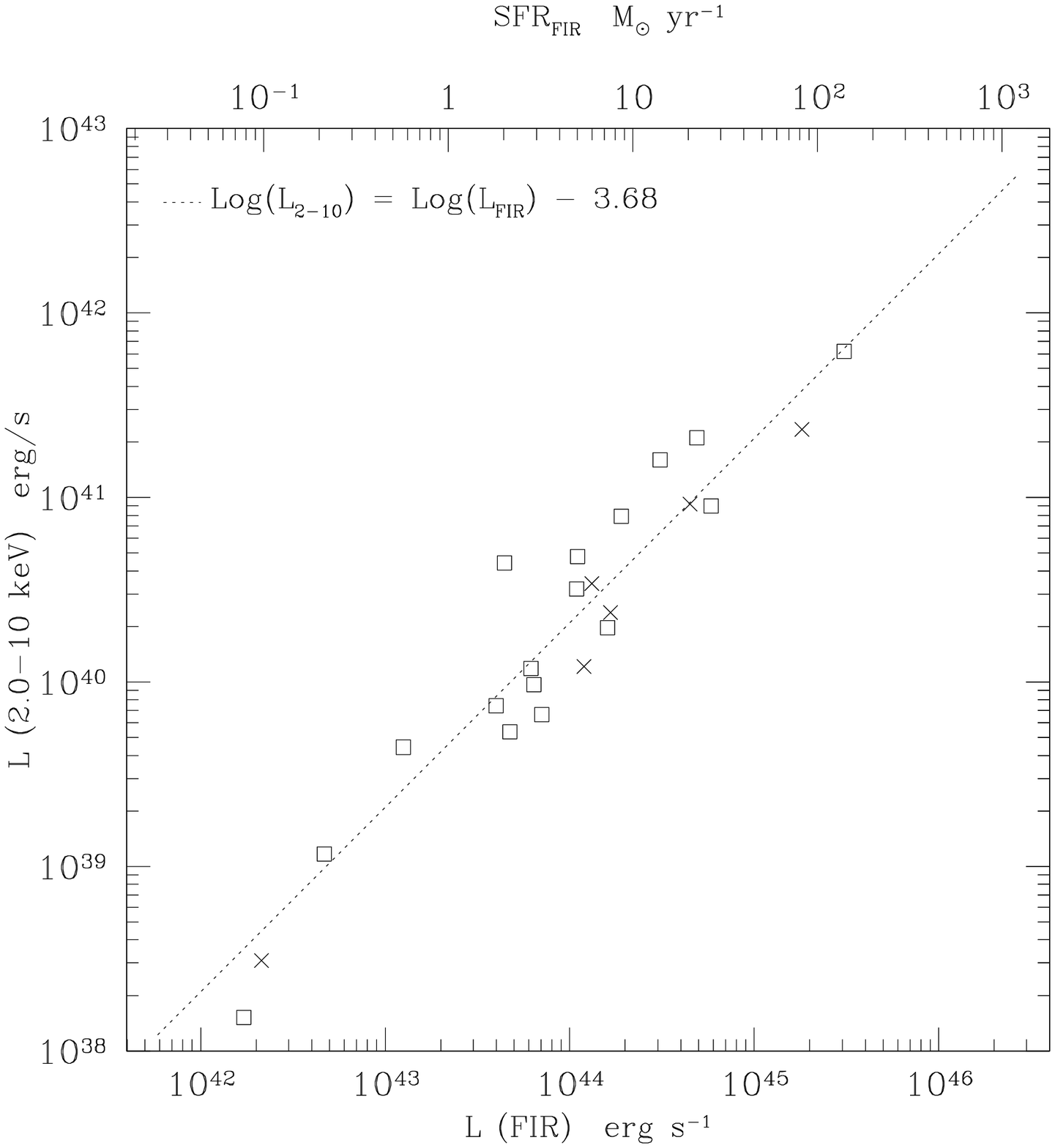}
      %\hskip.8cm
      \includegraphics[width=0.49\textwidth]{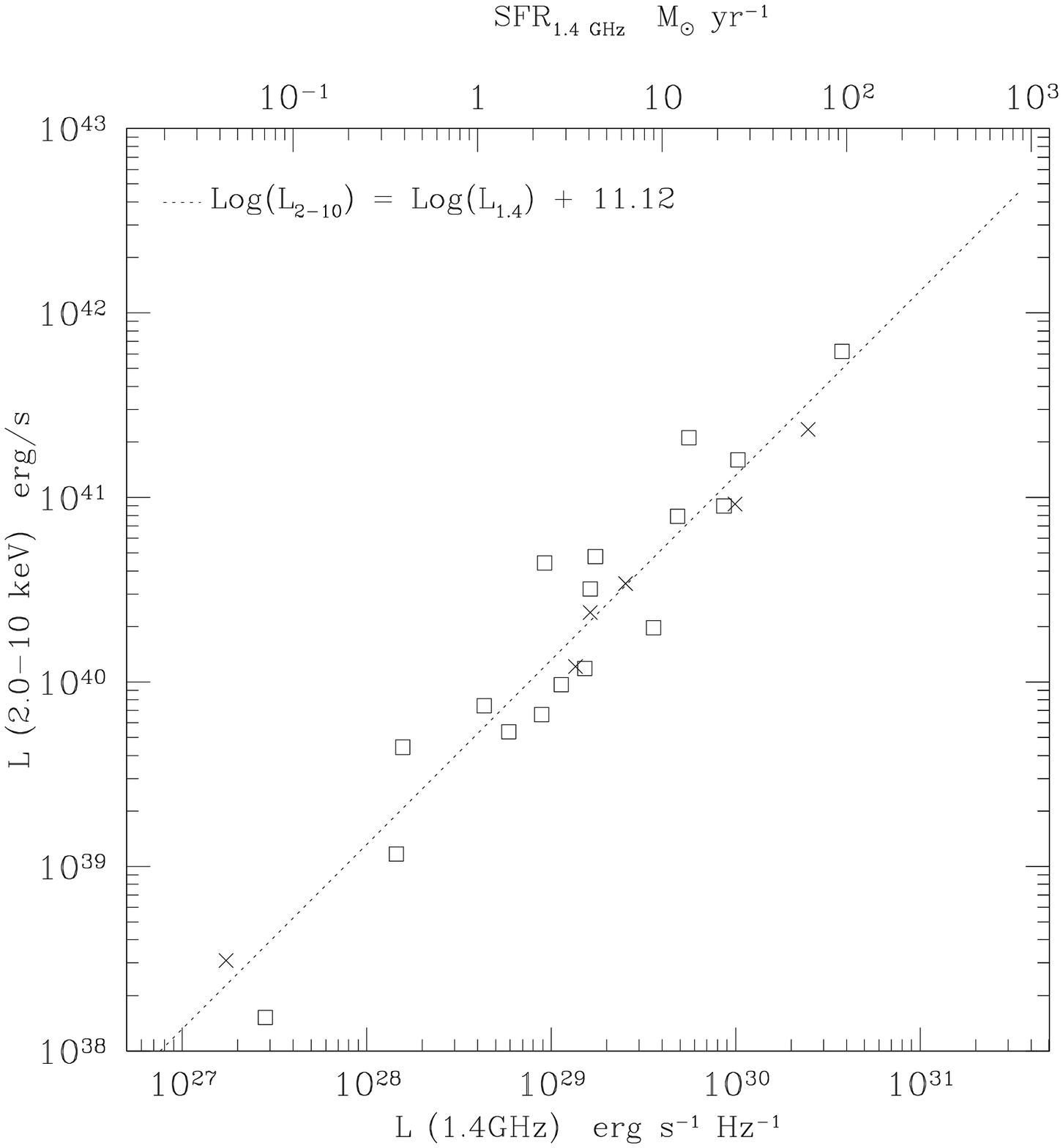}
  \end{center}
  \caption {The 2.0--10 keV luminosity of local star-forming galaxies
  vs.~radio and FIR ones. Symbols as in Fig.~(\ref{pranalli-E3_fig:fig1});
  dotted lines: Eqs.~(\ref{eq:firx2},\ref{eq:radiox2}).
  }
  \label{pranalli-E3_fig:fig2}
\end{figure*}

\subsection*{Hard X-rays}

In Fig.~(\ref{pranalli-E3_fig:fig2}) we plot 2--10 keV luminosities
versus FIR and radio ones.  Least-squares fits yield:\\
\begin{eqnarray}
\Log (L_{\rm 2-10}) &= &(1.08\pm 0.09) ~\Log (L_{\rm FIR}) 
	-7.1\pm 4.2 \label{pranalli-E3:bf210a} \\
\Log (L_{\rm 2-10}) &= &(1.08\pm 0.09) ~\Log (L_{\rm 1.4})
	 +8.8 \pm 2.7 \label{pranalli-E3:bf210b}
\end{eqnarray}\\
with $s\simeq 0.30$ and 0.29 respectively.  The linearity
and the dispersion are not significantly changed neither by the
inclusion of the supplementary sample ($L_{2-10}\propto L_{\rm
FIR}^{1.04\pm 0.07}$ and $L_{2-10}\propto L_{\rm 1.4}^{1.01\pm 0.07}$,
$s\simeq 0.27$ and 0.26 respectively), nor by the use of the $60\mu$
luminosity ($L_{2-10}\propto L_{60\mu}^{1.01\pm 0.07}$).

By assuming an exactly linear slope, the best fit relations 
for the local (local+supplementary) sample become:\\
\begin{eqnarray}
\Log (L_{\rm 2-10}) &= &\Log (L_{\rm FIR}) - 3.62~~ (3.68) \label{eq:firx2}\\
\Log (L_{\rm 2-10}) &= &\Log (L_{\rm 1.4}) + 11.13~~ (11.12) 
   \label{eq:radiox2}
\end{eqnarray}
with $s \simeq 0.29$ for both fits. There is no significant improvement
(less than $1\sigma$)
in the free-slope fits with respect to the linear slope ones.

\section{X-rays and the Star Formation Rate}      \label{biases_section}

The existence of a tight linear relation implies that the three
considered bands all carry the same information. Since the radio and
far infrared luminosities are indicators of the SFR, the 0.5--2 keV and
2--10 keV luminosities should also be SFR indicators.
However, before attempting to calibrate such relationships, we should    
consider the possible existence of selection effects.

\citet{hfs95} made a special effort in obtaining {\em nuclear} nebular
spectra, so that a reliable spectral classification of the central
engine could be derived.      The main concern is the possibility that
the \Hii galaxies in the HFS97 sample could host a Low Luminosity AGN
(LLAGN), which might significantly contribute to the overall energy
output. To check for this possibility, \citet{ulvestad} observed with
the VLA at 1.4 GHz a complete sample of 40 Sc galaxies in HFS97 with
\Hii spectra and did not find any compact luminous radio
core. Instead, they found that the radio powers and morphologies are
consistent with star formation processes rather than by
accretion onto massive black holes; thus they suggest that \Hii nuclei
intrinsically lack AGN. Therefore we believe that the
HFS97 classification is reliable and that our sample is not polluted
by AGN.

It is also worth noticing that the soft X-rays relationships may
involve some further uncertainties related to the possible presence of
intrinsic absorption (negligible in the 2--10 keV band for
column densities usually found in normal galaxies).  An example of
this effect is the southern nucleus of NGC\,3256 (see the Appendix), a
dusty luminous merger remnant with two bright radio-IR cores where
star formation is ongoing: while both of them fall on the
radio/hard X-ray relation, only the northern core is on the
radio/soft X-ray relation because the southern one lies behind a dust
lane which absorbes at all wavelengths from $\sim 1\mu$ to $\sim 2$
keV.  The quasi-linearity of the soft X-ray relations suggests that
absorption is unlikely to be relevant for the majority of the objects in
our sample; however this effect may become significant at
cosmological distances ($z\gtrsim 1-2$) where galaxies have more dust
and gas at their disposal to form stars.

Thus we feel confident to propose the use of X-ray luminosities
as SFR indicators. From eqs.(\ref{eq:firx},\ref{eq:radiox},\ref{eq:firx2},%
\ref{eq:radiox2}) we derive:
\begin{eqnarray}
{\rm SFR}~=&2.2\cdot10^{-40} ~L_{\rm 0.5-2 keV} &\qquad\mbox{M$\sun$/yr}\label{pranalli-E3:eqsfrsoftX2}\\
{\rm SFR}~=&2.0\cdot10^{-40} ~L_{\rm 2-10 keV}  &\qquad\mbox{M$\sun$/yr}.\label{pranalli-E3:eqsfrhardX2}
\end{eqnarray}

     We also notice that there is growing evidence that star formation could play
a major role even among those objects classified as LLAGN. The preliminary results of the
\chandra LLAGN survey \citep{ho2001} show that only about one third
of LLAGN have a compact nucleus dominating the X-ray emission, while
in the remaining objects off-nuclear sources and diffuse emission 
significantly contribute to the overall emission.

\begin{figure}[t]    % firx e radiox
  \begin{center} 
      \includegraphics[width=0.49\textwidth]{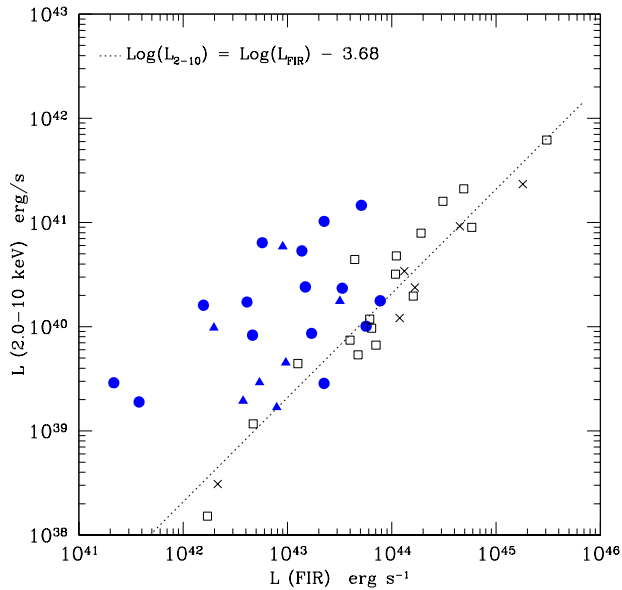}
  \end{center}
  \caption {2.0--10 keV vs.~FIR luminosities of star-forming galaxies
  and LLAGN. Filled triangles: LINERs, filled circles: Seyfert galaxies.
  Open symbols as in Fig.~(\ref{pranalli-E3_fig:fig1});
  line: Eq.~(\ref{eq:firx2}).
  }
  \label{figseyfirx}
\end{figure}

     Following this investigation, we have analyzed the relations
between radio/FIR/X-ray luminosities for the spiral galaxies in the
\citet{terashima} sample of LLAGN, drawn from HFS97 and observed with
ASCA, comprising 7 LINERs and 15 Seyfert's with $4\cdot 10^{39}\lesssim
L_{2-10}\lesssim 5\cdot 10^{41}$ erg s$^{-1}$.  We find that
the X-ray/FIR and X-ray/radio luminosity ratios generally exceed those
of star-forming galaxies, but about one third of the objects have ratios
falling on the same {\em locus} of the star-forming galaxies
(Fig.~\ref{figseyfirx}). Therefore, the nuclear X-ray emission of
these last LLAGN must be comparable to or weaker than the emission from
star formation related processes.  Moreover, the infrared (IRAS band)
colours of these objects are also similar to those of star-forming
galaxies, and completely different from those of QSOs,
thus suggesting that the FIR luminosities of LLAGN may be powered by
star formation.

\section{Star-forming galaxies in the Hubble Deep Field}
\label{hdf_section}

The 1~Ms {\em Chandra} \citep{bran01} and the radio (8.4 GHz:
\citealt{vla98}; 1.4 GHz: \citealt{vla00} and \citealt{wsrt})
catalogues of the Hubble Deep Field North (\object{HDFN}) reach a
limiting flux which is deep enough to detect star-forming galaxies at
redshifts up to $z\sim1.3$, and can be used to check whether the
radio/X-ray relation holds also for distant galaxies.

We searched for X-ray counterparts of radio sources in the the
\citet{vla98} catalogue which contains optical and IR identifications
allowing the selection of candidate star-forming galaxies.  Our
selection criterium has been to include all galaxies with Spiral or
Irregular morphologies, known redshifts and no AGN signatures in their
spectra (from \citealt{vla98} or \citealt{pranalli-E3:cohen}).

The mean positional uncertainties of both \chandra (for on-axis
sources) and the VLA are $\sim 0.3\arcsec$, which added in quadrature
give $\sim 0.5\arcsec$. Using this value as the encircling radius for
coordinate matching 5 galaxies are found. However, there are two
effects that can increase this value:
\begin{enumerate}
\item the shape and width of the \chandra PSF strongly depend on
the off-axis and azimutal angles. Since the 1 Ms HDFN data consist of 12
observations with different pointing directions and position angles,
there is no unique PSF model even for near on-axis sources;% \citet{bran01}
%estimates a positional uncertainty of $\sim 0.6\arcsec$ for on-axies sources
%in the HDFN. Adding this value to the VLA one gives $0.7\arcsec$, within which
%6 galaxies are matched;
\item a displacement of several kpc between the brightest
radio and X-ray positions, induced e.g.~by an ultraluminous X-ray binary
placed in a spiral arm and dominating the X-ray emission.
\end{enumerate}
Thus, by making cross-correlations with increasing encircling radii we found
that the number of coincidences increases up to a radius
of $1.0\arcsec$ (which yields 7 matchings). There are no further coincidences
up to a radius of several arcsecs, indicating that the sample should not
be contamined by chance coincidences. The 7 selected objects are 
listed in Table \ref{deepnomi}. 
Fluxes at 1.4\,GHz and spectral slopes were retrieved from
\citet{vla00} in 6 cases, and from \citet{wsrt} in one case.

\begin{table}  
\centering\begin{tabular}{cc }
{\sc Chandra} &{\sc VLA} \\
\hline
\object{CXOHDFN J123634.4+621212}   (134) &3634+1212\\
\object{CXOHDFN J123634.5+621241}   (136) &3634+1240\\
\object{CXOHDFN J123637.0+621134}   (148) &3637+1135\\
\object{CXOHDFN J123651.1+621030}   (188) &3651+1030\\
\object{CXOHDFN J123653.4+621139}   (194) &3653+1139\\
\object{CXOHDFN J123708.3+621055}   (246) &3708+1056\\
\object{CXOHDFN J123716.3+621512}   (278) &3716+1512\\
\hline
\end{tabular}
\caption{Identification of candidate star-forming galaxies in the \citet{vla98}
catalogues with their X-ray counterpart \citep{bran01} . The entry number in the
\chandra catalogue is shown in parenthesis.
\label{deepnomi}
}
\end{table}

\begin{figure}[]    % deep field hardness ratios
  \begin{center} 
     \includegraphics[width=0.49\textwidth]{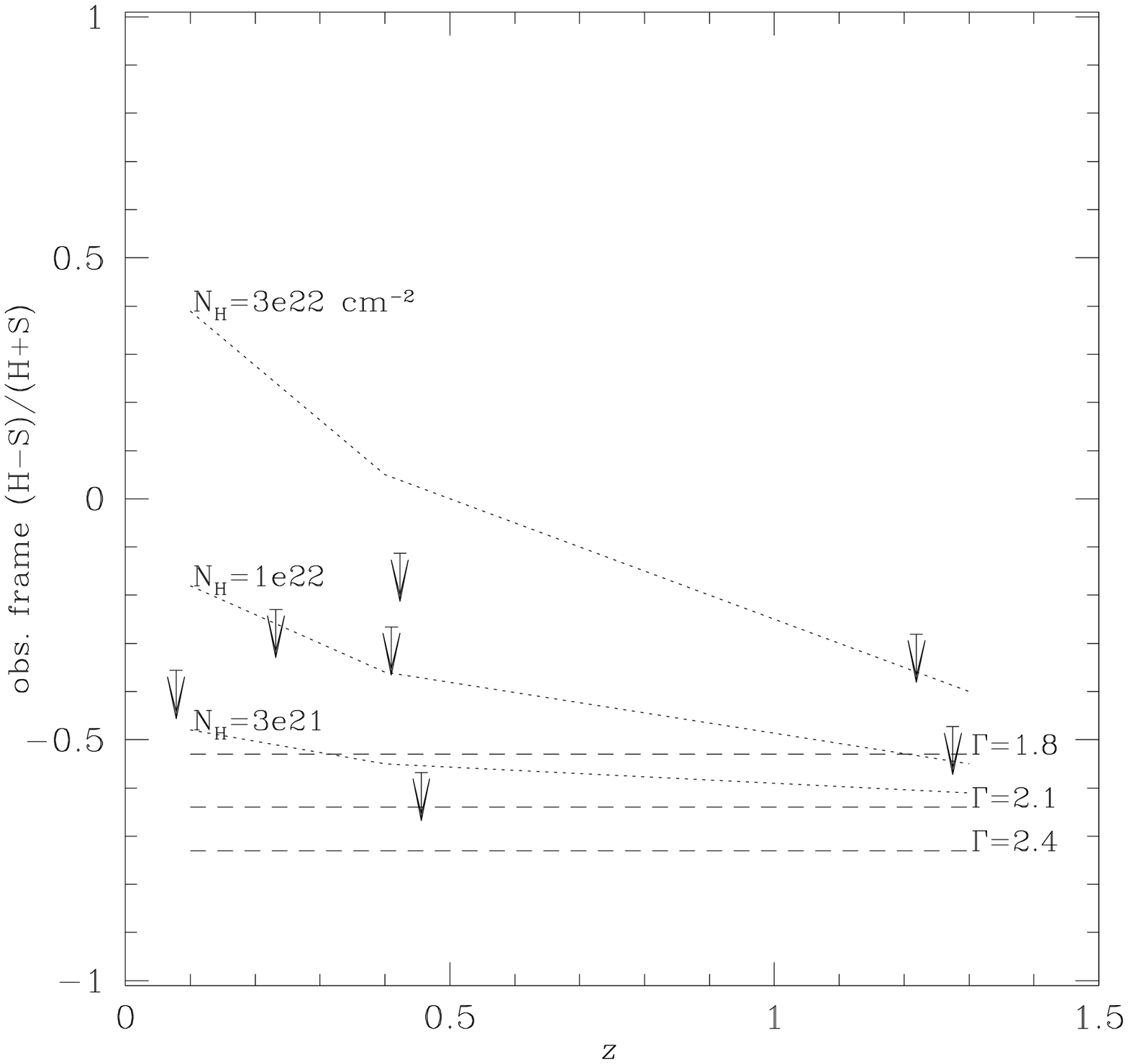}
  \end{center}
  \caption { Observer's-frame hardness ratios for galaxies in the HDFN;
there are only upper limits since none is detected in the 2--8 keV band
\citep{bran01}. 
The dotted lines show the {\em loci} for power-law spectra with slope $\Gamma=2.1$ and
different intrinsic rest-frame absorption; the dashed lines are for spectra of different
slopes and no intrinsic absorption.
\label{fig_hr}
}
\end{figure}

While rest-frame 0.5--2.0 and 2.0--10 keV luminosities could be
obtained by K-correcting the observer's frame counts, this would imply
the assumption of a spectral shape; but none of the seven deep field
galaxies is individually detected in the hard band, so that any
constraint on their spectra obtained with the use of a hardness ratio
diagram is too loose to be significant (Fig.~\ref{fig_hr}).
%Thus, the derived luminosities would suffer of an additional uncertainty
%coming from a somewhat arbitrary assumption of the spectrum.
However this problem may be partially circumvented by resizing the
X-ray bands in the observer's frame according to the redshift of the objects.
%However, it is possible to get rid of the K-correction if the redshift
%information is properly used: the bands from which counts are
%extracted can be resized according [...]    Thus
%the counts extracted with this method are those which were originally
%emitted in the rest-frame 0.5--2.0 and 2--8 keV%
It is therefore possible to give better constraints to the
spectrum of the deep field galaxies and derive better estimates of
their luminosities.

% ~~~~~~~~~~~~~~~~~~~~~~~~~~~~~
%  TABELLA: CHANDRA DEEP FIELD
% ~~~~~~~~~~~~~~~~~~~~~~~~~~~~~
\begin{table*}%[p]
\centering\begin{tabular}{ccccccccccccc}
\multicolumn{12}{c}{\sc Fluxes and Luminosities: Deep Sample} \\
\\
& &\multicolumn{3}{c}{\sc Soft X-rays}&\multicolumn{3}{c}{\sc Hard X-rays}&&\multicolumn{3}{c}{\sc Radio}\\
& &\multicolumn{3}{c}{\downbracefill}&\multicolumn{3}{c}{\downbracefill}&&\multicolumn{3}{c}{\downbracefill}\\
{\sc Source} &$z$ &{\sc Counts} &$F$ &$L$ &{\sc Counts} &$F$ &$L$ &$\Gamma$ &$F$ &$L$ &$\alpha$  \\
\hline
134 &0.456 &$31\pm  6$   &23   &29   &$22 \pm  6$  &28  &36   &$2.0_{-0.3}^{+0.4}$ & 210 &2.6  & 0.7\\%
136 &1.219 &$5.9\pm 3$   &6.7  &95   &$22 \pm  5$  &19  &260  &$1.6_{-0.5}^{+0.9}$ & 180 &26   & 0.7\\%
148 &0.078 &$14\pm  4$   &8.4  &0.23 &$< 2.1$     &$<4.1$ &$<0.11$ &$>2.6$ & 96  &0.027& 0.6 \\%
188 &0.410 &$15\pm  4$   &13   &13   &$ 5.2 \pm  4$  &5.8 &5.8  &$2.7_{-0.6}^{+1.5}$ &  83 &0.82 & 0.6\\%
194 &1.275 &$< 1.6$      &$<0.88$ &$<$14 &$23 \pm  5$  &20  &310  &$<0$&  60 &9.6  & 0.9\\%
246 &0.423 &$ 6.2\pm 4$&3.7  &4.0  &$ 5.4 \pm  4$  &7.5 &8.1  &$1.7_{-0.6}^{+1.4}$ &  36 &0.39 & 0.4\\%
278 &0.232 &$11\pm  4$ &5.8  &1.6  &$ 7.8 \pm  4$  &16  &4.4  &$1.5_{-0.4}^{+0.7}$ & 160 &0.44 & 0.2\\%
\hline
\end{tabular}
%\object{NGC253} &1 &2 &3 &4 &5 &6 &7 &8 &9 &10\\
%\object{M82} &1 &2 &3 &4 &5 &6 &7 &8 &9 &10\\
%\multicolumn{3}{c}{\sc Local Sample}\\ 
%\hline\hline
%
%\object{M82}*     &\object{NGC\,2276} &\object{NGC\,4449}\\
%\object{M101}    &\object{NGC\,2403} &\object{NGC\,4631}\\
%\object{M108}    &\object{NGC\,2903} &\object{NGC\,4654}\\ 
%\object{NGG\,891}  &\object{NGC\,3310}&\object{NGC\,6946}\\
%\object{NGC\,1569} &\object{NGC\,3367} &\object{IC\,342}  \\
%\object{NGC\,2146} &\object{NGC\,3690} \\
%\\
%\multicolumn{3}{c}{\sc Supplementary Sample }\\
%\hline\hline
%\object{NGC\,55}	&\object{NGC\,1672} &\object{NGC\,3256}\\
%\object{NGC\,253}*	&\object{NGC\,1808} &\object{Antennae}\\
%\end{tabular}
%{\sc VLA98 Id.} &{\sc Chandra \#} &$z$ &{\sc Extr. Band} &{\sc Soft Counts} &{\sc Hard Counts} &L210 &F210 &L210 &S14 &L14 &{\sc Notes} \\
%4,6  22:30 -- rit: 8.6 la fra frp lor -- albergo: 50eur bb 85 mp 100 pc
%CXOHDFN J123634.4+621212
%CXOHDFN J123634.5+621241
%CXOHDFN J123637.0+621134
%CXOHDFN J123651.1+621030
%CXOHDFN J123653.4+621139
%CXOHDFN J123708.3+621055
%CXOHDFN J123716.3+621512
\caption{Data for deep field galaxies (counts, fluxes, luminosities). 
 $\Gamma$ is the
 best fit X-ray slope (photon index), $\alpha$ the radio slope (energy index).
 Sources are identified via their entry number
 in the \citet{bran01} catalogue (cfr.~Table \ref{deepnomi}).
 X-ray fluxes in $10^{-17}$
  erg s$^{-1}$ cm$^{-2}$, radio fluxes in $\mu$Jy. X-ray luminosities in
 $10^{40}$ erg s$^{-1}$, radio luminosities in $10^{30}$ erg s$^{-1}$ Hz$^{-1}$. 
 X-ray counts are extracted in redshifted bands (soft band $:=0.5$--$2.0/(1+z)$ keV,
 hard band $:=2.0/(1+z)$--$8/(1+z)$ keV). X-ray fluxes and luminosities are in rest-frame
 0.5--2.0 and 2.0--10 keV bands, radio ones at rest-frame 1.4\,GHz.
 For source \#194, the absorbed flux and luminosity are quoted. The unabsorbed luminosity
 is $\lesssim 4\cdot 10^{42}$ erg s$^{-1}$.
\label{tab_deep}
}
\end{table*}

%In order to obtain information about their spectra, for each source
%we extracted counts in {\em redshifted} bands corresponding to
%rest-frame 0.5--2.0 and 2.0--8.0 keV. The advantage is that for $z\sim
%1$ the hard band becomes $\sim$1-4 keV in observer's frame, which
%corresponds to the range of highest sensitivity of {\em Chandra}.
Thus we redefine the soft and hard%
\footnote{Since \chandra has very poor sensitivity between 8 and 10
keV, the use of the reduced 2--8 keV band enhances the signal/noise
ratio. Note that while counts are extracted in the 2--8 keV band,
fluxes and luminosities are always extrapolated to the 2--10 keV
band.} bands as %0.5--$2.0/(1+z)$ and
%$2/(1+z)$--$8/(1+z)$
the [0.5; $2.0/(1+z)$] and [$2.0/(1+z)$; $10/(1+z)$] intervals,
respectively.  Another advantage of this procedure is that the higher
the redshift, the more akin the new hard band is to the zone of
maximum sensitivity of \chandra ($\sim1-4$ keV). Note that since the
ACIS-I detector has almost no sensitivity below 0.5 keV, we fixed this
energy as the lower limit for count extraction. For the two highest
redshift galaxies, this reduces the soft band to 0.5--0.9 keV, still
significantly larger than the ACIS-I energy resolution (FWHM~$\lesssim
100$ eV).

We extracted counts in circular regions around our selected targets
(radius 5 pixels); background was taken in annuli surrounding the
targets.  The net counts did not show dependence on the choice of
background regions; they were converted in count rates with the
exposure times listed in the \citet{bran01} catalogue.  Counts and
rates are reported in Table \ref{tab_deep}.  Best-fit slopes
reproducing the soft/hard count ratio were derived by assuming a
power-law spectrum with Galactic absorption. We find that six objects
have spectral slopes falling in the range $\Gamma\sim 1.5$--$2.7$
(Table \ref{tab_deep}).  To check whether these spectra are consistent
with those of the galaxies in the local sample we calculated the
observed soft/hard flux ratio for galaxies in the local sample: the
median value for this flux ratio is 0.95 leading to a slope
$\Gamma\sim 2.1$.  The count ratios for each of the six deep field
galaxies are consistent within 1--2$\sigma$ with the $\Gamma\sim 2.1$
slope.  The remaining source (\#194 in Table \ref{tab_deep}, at
$z=1.275$) has an upper limit on the soft X-ray counts and a count
ratio not consistent at the $5\sigma$ level with the unabsorbed
$\Gamma\sim 2.1$ spectrum: it requires an inverted spectrum
($\Gamma<0.1$) if no absorption is assumed, otherwise, if we assume
$\Gamma=2.1$, the intrinsic absorbing column has to be $N_{\rm
H}\gtrsim 3.4\cdot 10^{22}$ cm$^{-2}$.

\begin{figure*}[t]    % deep field lumin
  \begin{center} 
      \includegraphics[width=0.49\textwidth]{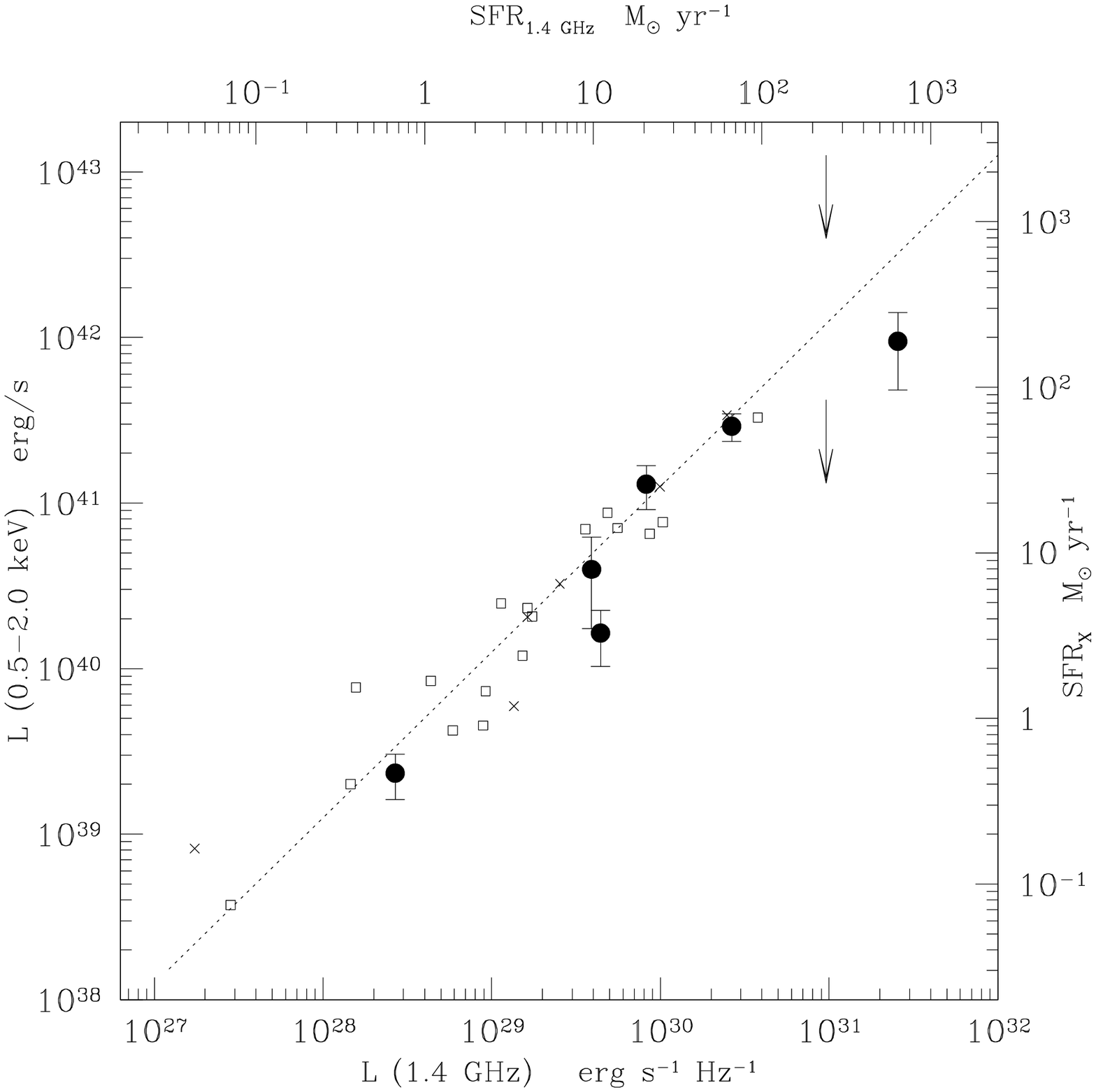}
      %\hskip.8cm
      \includegraphics[width=0.49\textwidth]{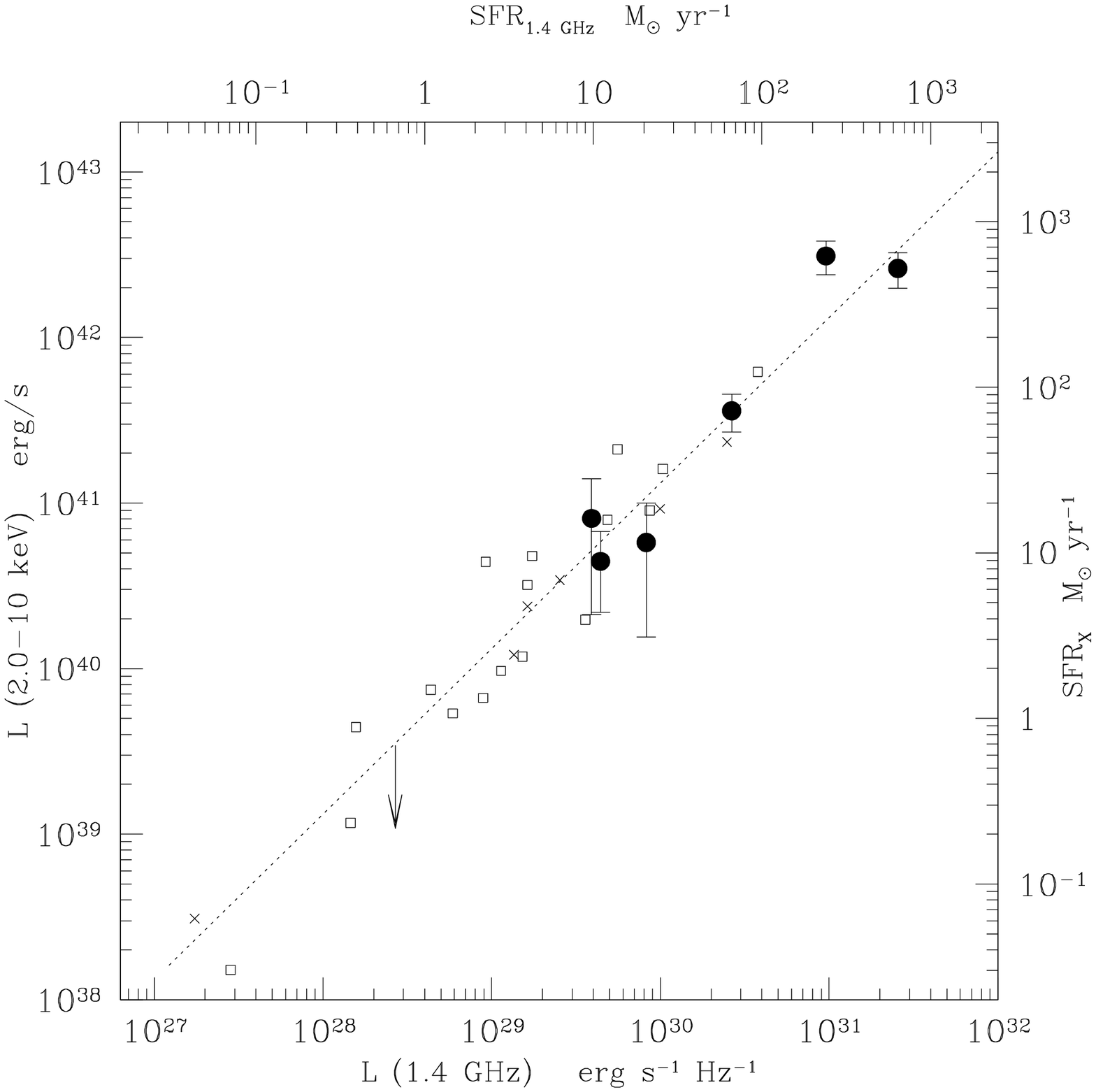}
  \end{center}
  \caption { The radio/X-ray luminosity relation for galaxies in the Hubble 
   Deep Field (filled circles). Open squares and crosses: local galaxies
    as in Fig.~(\ref{pranalli-E3_fig:fig1}); dotted lines:
    linear fits for local galaxies (Eq.~% 
   \ref{eq:radiox}). The two most luminous galaxies
   are at $z=1.275$ and 1.219 in order of increasing radio luminosity.
    The two upper limits (3$\sigma$ level) in the left panel refer to source \#194 (see text) 
  and show the unabsorbed luminosity as estimated with $\Gamma\sim 0.1$ and no intrinsic
   absorption (lower) and with $\Gamma\sim 2.1$ and $N_{\rm H}\sim 3.4
   \cdot 10^{22}$ cm$^{-1}$ (upper).
  \label{fig_deeplumin}
  }
\end{figure*}

With the best-fit slopes we derived soft and hard band fluxes and
luminosities (Fig.~\ref{fig_deeplumin}).
The linear radio/X-ray correlations hold also for the Deep Field galaxies;
the dispersion of the relations given in \S\ref{correlazioni_section}
is not changed by the inclusion of the deep field objects.
A similar relation was also
found in a recent paper by \citet{bauer02}.
%
%: the best-fit
%relations for the deep field sample are:
%\begin{eqnarray}
%\Log (L_{\rm 0.5-2}) &= (0.9\pm 0.1) &\Log (L_{\rm 1.4}) + 14\pm 3 \\
%\Log (L_{\rm 2-10}) &= (1.0\pm 0.1) &\Log (L_{\rm 1.4}) + 11\pm 2.98 
%   \label{eq:radioxhdf}\qquad. 
%\end{eqnarray}
%Note that in these fits the two ($3\sigma$) upperlimits were treated
%as detections. 
%the slope of the soft X-ray/radio relation is still
%consistent with unity within $1\sigma$, despite 

%Although the presence of one absorbed source does not alter the properties of
%the radio/soft X-ray relation, its presence in the sample
%. Deeper \chandra observations providing larger samples
%and better statistics will allow a better constraint on the presence
%of intrinsic absorption in high redshift galaxies.

\section{X-ray number counts and background}     \label{fondox_section}

An estimate of the contribution of star-forming galaxies to the cosmic
X-ray background (XRB) has been attempted several times
(e.g. \citealt{bookbinder80}, \citealt{gp90}, \citealt{moran99}).  The
main purpose for the earlier studies was the possibility to explain
the flatness of the XRB spectrum via the X-ray binaries powering the
X-ray emission of these galaxies. Although AGN have since been
recognized to provide by far the most important contribution to the
XRB \citep{sw89,comastri95}, the ongoing deep \chandra and \XMM
surveys offer unique opportunities to both test the AGN models and pin
down the contribution from other kind of sources.  Here, making use of
the X-ray/radio relationships of \S\ref{correlazioni_section}, we
perform an estimate of the contribution to the XRB by star-forming
galaxies.

We consider the radio sub-mJy population associated with faint blue
galaxies at high redshifts ($22\lesssim V\lesssim 27$, $0.5\lesssim
z\lesssim 1.5$; \citealt{windhorst90}) representing an early era of
star formation in the universe \citep{haarsma00}. This strongly
evolving population accounts for the majority of the number counts
below $\sim 0.5$ mJy \citep{windhorst85} and contribute about half of
the radio cosmic background at 1.4 GHz \citep{haarsma00}.  The deepest
radio surveys have been performed at 1.4 GHz \citep{vla00}, 5 GHz
\citep{fomalont91} and 8.4 GHz \citep{vla98}.  Although a quick
estimate of the contribution to the XRB may be worked out by simply
applying Eqs.~(\ref{eq:radiox},\ref{eq:radiox2}) to the total radio
fluxes obtained by integrating over the deepest radio \lognlogsa, it
is instructive to first derive the X-ray source counts and compare
them with the deepest counts obtained by {\em Chandra}.  In order to
derive the X-ray counts from the radio ones a full knowledge of the
redshift distribution and spectra of the sources would be
required. Under the simplifying assumption that the sub-mJy population
lies at a redshift $\bar{z}$, so that the K-correction term is the
same for all sources, the differential counts are obtained as:
\begin{equation}
n(S_{\rm X})=n(S_{1.4}\cdot \kappa^{\rm X}_{1.4})
\end{equation}
where $n(S)$ are the differential number counts, and $\kappa^{\rm
X}_{1.4}=\kappa^{\rm X}_{1.4}(\bar{z},\alpha,\Gamma)$ is the X-ray
band/1.4 GHz luminosity ratio (eqs.\ref{eq:radiox},\ref{eq:radiox2})
which depends on the redshift and on the radio and X-ray spectral indices
via the K-correction:
\begin{equation}
\kappa^{\rm X}_{1.4}(\bar{z},\alpha,\Gamma) =
\left.\kappa^{\rm X}_{1.4}\right|_{z=0} \cdot (1+\bar{z})^{-(\Gamma-1)+\alpha}\qquad.
\end{equation}

Concerning the sub-mJy \lognlogsa, \citet{vla00} gives $n(S)=(2.51\pm
0.13)\cdot 10^{-3} S^{-2.38\pm 0.13}$ deg$^{-2}$ Jy$^{-1}$ as a
best-fit to the differential number counts at 1.4 GHz in the range
45--1000, while \citet{fomalont91} find $n(S)=1.2\cdot 10^{-3}
S^{-2.18\pm 0.19}$ at 5 GHZ in the range 16--1500 $\mu$Jy. The number
density at 4 $\mu$Jy (as estimated from fluctuation analysis) is
consistent with extrapolation of the 16--1500 $\mu$Jy slope.

The mean radio spectral index is in the range $\alpha=0.3$--0.7.
\citet{fomalont91} report their distribution as having a moda of
$\alpha=0.5$, a median of $\alpha=0.38$, and an average of
$\alpha=0.28$ (indices measured between 1.5 and 5 GHz). From
\citet{vla00} data we find an average $\alpha=0.47$ (between 1.4 and
8.4 GHz), when we consider detections at both frequencies, and
$\alpha=0.67$ when we consider only detections at 1.4 GHz and treat
upper-limits at 8.4 GHz as detections. The latter, steeper slope is
also more consistent with the average index of our HDFN sample (Table
\ref{tab_deep}).  Here we assume $\alpha=0.5$, and estimate the
uncertainty in the K-correction due to the radio spectral index to be
around a $15-20\%$.  

     According to the results of \S\ref{hdf_section}, we further assume an
average X-ray spectral slope $\Gamma = 2.1$. For the sake of simplicity
we also assume that the objects are placed at $\bar{z}\sim 1$, the mean redshift
for the sub-mJy galaxies which, according to \citet{windhorst90},
are distributed in the redshift interval 0.5--1.5 with a peak at
$z\sim 1$. To estimate the effect on the counts due to the actual distribution
of the sources in this redshift interval we consider a simplified case
in which they are equally distributed at redshifts $z=0.5$, 1.0 and 1.5. Since
the effect enters in the computations via the K-correction term, we find
that the predicted number counts would increase by only $\sim 10\%$. As a
last remark we notice that a fraction, as yet undefined, of the X-ray
spectra might steepen at energies $\gtrsim 10$ keV, thus entailing a decrease in the
predicted source counts; however, given the redshift range under consideration,
the predicted soft X-ray source counts should not be affected.

%%%%%%%%%%%%%%%%%%%%%%%%%%%%%%%%%%%%%%%%%%%%%%%%%%%
% FIGURA: log N log S
%%%%%%%%%%%%%%%%%%%%%%%%%%%%%%%%%%%%%%%%%%%%%%%%%%%
\begin{figure*}[t]    % log N -- log S
  \begin{center}
  \includegraphics[width=0.49\textwidth]{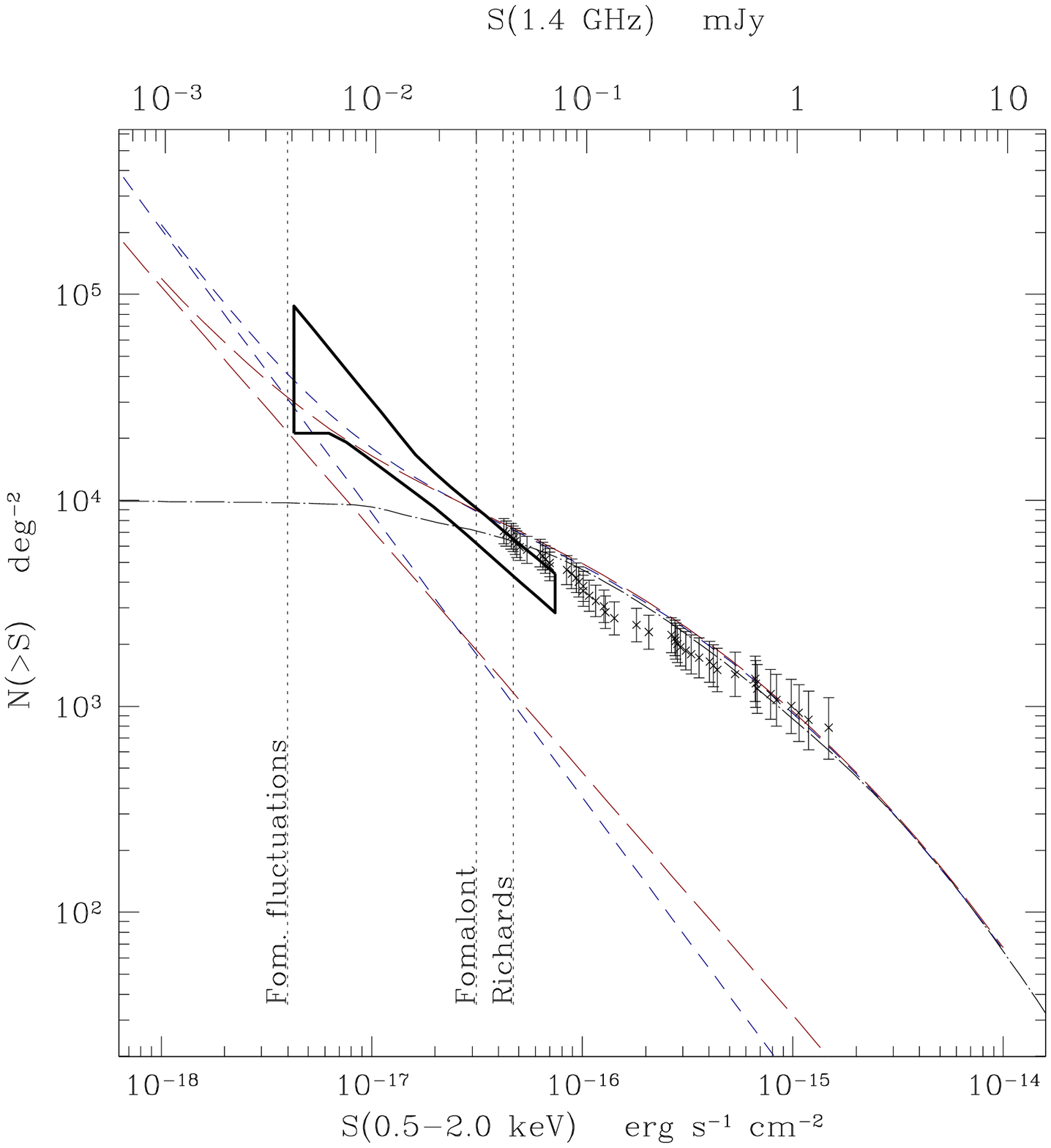}
  %\hskip.8cm 
  \includegraphics[width=0.49\textwidth]{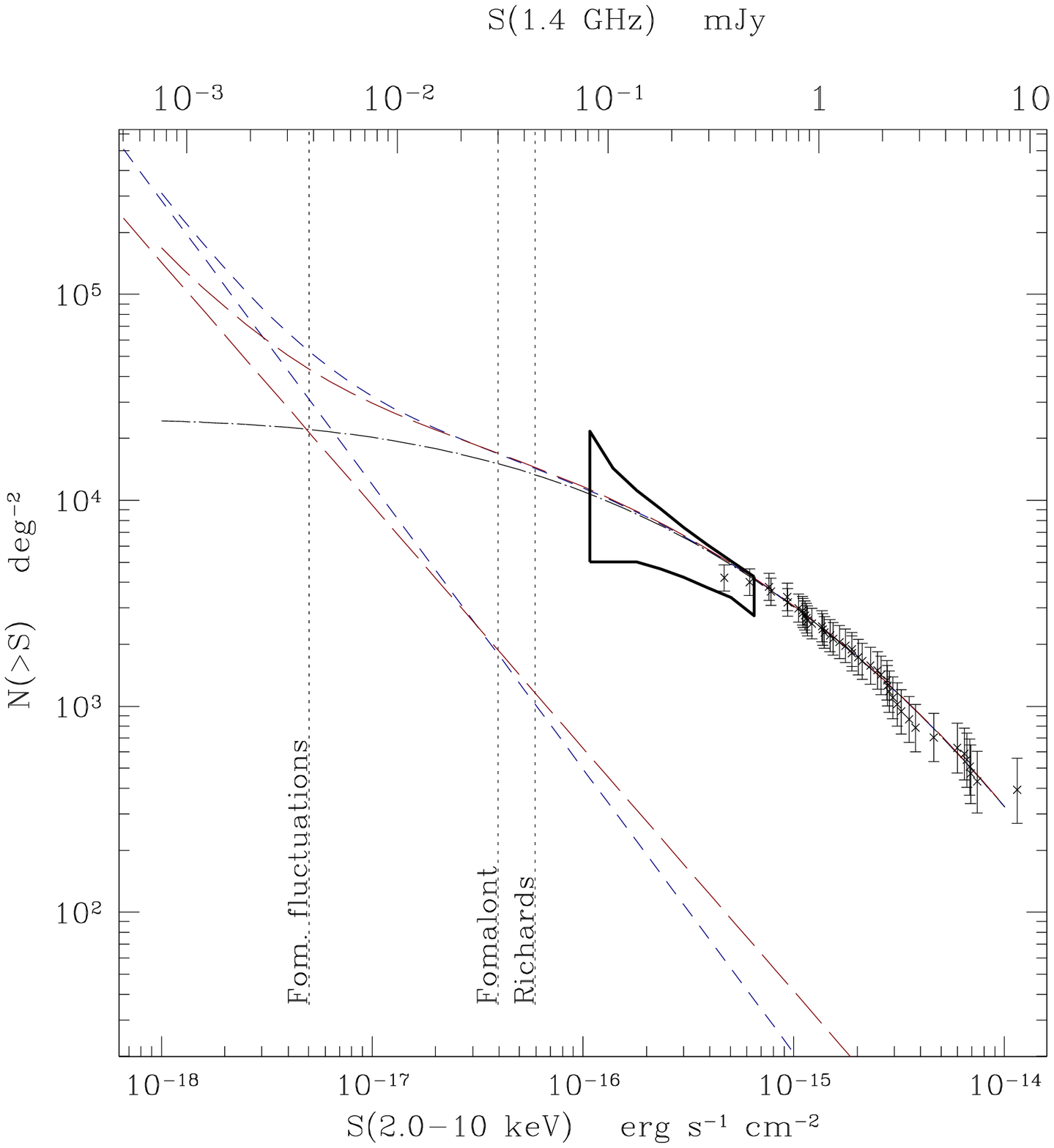}
  \end{center} 
  \caption { X-ray counts derived from deep radio \lognlogsa.
    The short-dashed lines
  represent the 1.4 GHz \lognlogs \citep{vla00} with %(lower line) and without (upper line)
  K-corrections assuming $\bar{z}=1$, $\Gamma=2.1$ and
  $\alpha=0.5$.  The long-dashed lines represent the
  \lognlogs by \citet{fomalont91} reported at 1.4 GHz with $\alpha=0.5$
  (K-corrections like for the 1.4 GHz
  one). The dots are the observed X-ray number counts in the HDFN
  \citep{bran01}; while the horn-shaped symbols show the results from X-ray fluctuation
  analysis \citep{miyaji02a,miyaji02b}. The dot-dashed line 
  shows the number counts from AGN synthesis models
  \citep{comastri95}.  The vertical dotted lines show the limiting
  sensitivities for the radio surveys. The total of AGN plus galaxy counts
  predicted from the radio counts of \citet{vla00} and \citet{fomalont91} 
  are also shown with short and long dashed lines, respectively.
  \label{fig_lognlogs} }
\end{figure*}

The X-ray number counts predicted 
with the above assumptions (i.e.~$\alpha=0.5$, $\Gamma=2.1$)
from \citet{vla00} and
\citet{fomalont91} \lognlogs are shown in Fig.~(\ref{fig_lognlogs}),
along with observed counts from \citet{bran01}, limits from
fluctuation analysis \citep{miyaji02a,miyaji02b} and predicted counts
from AGN synthesis models \citep{comastri95}. The latter are
essentially in agreement with the results of more recent synthesis
models \citep{gilli01}.  The extrapolations of the radio \lognlogs
below $\sim 50 \mu$Jy and down to a few $\mu$Jy at 1.4 GHz ($\sim
6\cdot 10^{-18}$ erg s$^{-1}$ cm$^{-2}$ in the 0.5--2.0 keV band) are
consistent with the limits from fluctuation analysis \citep{miyaji02b}
and do not exceed the X-ray expected number counts.  Since
\citet{miyaji02b} suggested the emergence of a new population (beyond
that of AGN which dominates at brighter fluxes) in the 0.5--2.0 keV
band at fluxes around $10^{-17}$ erg s$^{-1}$ cm$^{-2}$, it is
tempting to identify this new population with the sub-mJy galaxies. It
should be noted, however, that eqs.(\ref{eq:radiox},\ref{eq:radiox2})
may not apply to the entire sub-mJy population. For instance, a
fraction (up to a $15-20\%$, \citealt{haarsma00}) of the sources may
still present a dominant contribution from an AGN. It follows that the
derived \lognlogs should be regarded as an upper limit to the X-ray
counts from star-forming galaxies.

     As a final remark it should be pointed out that our results depend on 
the basic assumption of a strict linearity between the radio and X-ray 
luminosities. Had we assumed a non-linear relationship, such as Eq.~(\ref{eq:radiox1}) 
for the soft X-rays or Eq.~(\ref{pranalli-E3:bf210b}) for the 2-10 keV band, we would
have found an increase or decrease, respectively, of about $50\%$ in the predicted 
counts at a flux level around $10^{-17}$ erg s$^{-1}$ cm$^{-2}$. A larger, well defined 
sample of star-forming galaxies down to the faintest radio and X-ray fluxes 
would obviously be of great importance to better constrain the X-ray vs.~radio 
luminosities relationship. It might be argued that a more direct derivation 
of the source counts could be made by adopting the observed fluxes, rather 
than the luminosities of the objects. However, this procedure would entail 
an arbitrary extrapolation of the X-ray/radio flux ratio for almost two 
orders of magnitude at fluxes fainter than the \chandra deep field. On the 
other hand, the radio luminosity interval of Fig.~(\ref{fig_deeplumin})
 essentially encompasses 
the radio power of the sub-mJy population if placed in the redshift interval
mentioned above.

The derived X-ray number counts can be integrated to estimate
the contribution to the XRB. For the observed 2--10 keV background we
take the \XMM value of $2.15\cdot 10^{-11}$ erg s$^{-1}$ cm$^{-2}$
\citep{lumb02}, which is comprised between the ASCA \citep{gendreau95}
and BeppoSAX \citep{vecchi99} figures.  The integration of counts
derived from the 1.4 GHz \lognlogs \citep{vla00}, performed in its
validity range ($5.9\cdot 10^{-17}-1.3\cdot 10^{-15}$ erg s$^{-1}$
cm$^{-2}$, corresponding to 45-1000 $\mu$Jy at 1.4 GHz), yields a
contribution to the XRB of $4.0\cdot 10^{-13}$ erg cm$^{-2}$ s$^{-1}$
($\lesssim 2\%$ of the observed background). By extrapolating to
$10^{-18}$ erg s$^{-1}$ cm$^{-2}$ ($\sim 1 \mu$Jy) the
contribution would increase to $2.3\cdot 10^{-12}$ erg cm$^{-2}$ s$^{-1}$
($11\%$).  Integration of counts from the flatter
\citet{fomalont91} \lognlogs in the 1-1000 $\mu$Jy range yields a
contribution of $1.4\cdot 10^{-12}$ erg cm$^{-2}$ s$^{-1}$
($6.4\%$). We also note that 1 $\mu$Jy at 1.4 GHz is a limit $\sim
3-5$ times fainter than the constraint from radio fluctuation
analysis; it is unlikely that the radio \lognlogs could sustain its
slope below this limit, otherwise the integrated radio emission from
weak sources would diverge \citep{fomalont91}.

\section{Conclusions}

%The basic ideas in the interpretation of the radio/FIR relation lie
%in the mechanisms of star formation and evolution, leading to cosmic
%ray acceleration and dust heating.  Our results suggest that the 2--10
%keV X-ray emission of galaxies is closely correlated with the radio
%and far infrared ones. The origin of the hard X-ray emission must thus
%be closely linked to star formation too. There are a few
%possibilities, i.e.  X-ray binaries, stellar mass black holes, hot
%($\sim 6$ keV, see \cite{pranalli-E3:cappi}) plasma, diffuse
%emission; however a basic picture could be sketched. Massive stars
%which heat the dust by means of their UV radiation, also end up with a
%supernova explosion in which a remnant and a compact object are
%formed. The remnant could be the site of cosmic ray acceleration and
%plasma heating. Compact objects which were in a binary system before
%the explosion can become X-ray binaries if the bynary system has
%survived. A population of X-ray binaries can thus be
%formed, which number should be proportional to the SFR.

We have analyzed a small, but well defined sample of 17 star-forming
galaxies, extracted from the HFS97 catalogue, for which there is a
homogeneous information on optical, FIR, radio and X-ray bands (local
sample). In agreement with previous work \citep{djf92} we find that
the logarithms of the soft (0.5--2 keV) X-ray luminosities (corrected
for Galactic absorption only) are linearly correlated with the
logarithms of both radio (1.4 GHz) and FIR luminosities.  We have
extended our analysis to the harder X-ray band, essentially free from
internal absorption which may affect the soft X-ray fluxes, and found
that there is a tight linear correlation between the X-ray
luminosities in the 2--10 keV interval with both the radio and the FIR
luminosities, normally assumed as the indicators of the star formation
rate. The addition of 6 galaxies (supplementary sample) homogeneous
with, but not included in HFS97, does not modify these results.
We conclude that the origin of the hard X-ray emission must be
closely related to star formation and calibrate an X-ray SFR
indicator.

Candidate starburst galaxies have been selected in the HDF North, with
redshifts up to $z\,\sim\,1.3$, and their rest-frame X-ray
luminosities are computed by extracting counts in redshifted bands
from the \chandra observation of the HDFN. With this approach we have
shown that the 2--10 keV/radio linear correlation holds up to $z\sim
1.3$, encompassing five orders of magnitude in luminosity, up to
$L_{2-10} \sim$ several $10^{42}$ erg~s$^{-1}$ and a corresponding
star formation rate $\sim 1000~ {\rm M}\sun$ yr$^{-1}$.  The fit to
the 0.5--2.0 keV/radio data is also linear up to $z\sim 1.3$. However,
the count ratio for the highest redshift galaxy at $z=1.275$ requires
significant absorption if a spectral slope of $\Gamma\sim 2.1$ is
assumed; thus this galaxy shows, on a larger scale, the same behaviour
as that of the southern core in NGC\,3256 (see the Appendix).
Therefore, while the linearity of the relations involving soft X-ray
luminosities remains statistically significant, at high redshift
(where galaxies are supposed to have more gas at their disposal to
form stars, and so their X-ray emission is more likely to be absorbed)
the 2--10 keV luminosity is a more secure indicator of the SFR.

     As an additional investigation we have also analyzed a sample of LLAGN
(LINERs and Seyfert's) included in HFS97 \citep{terashima}:
while, as expected, the X-ray luminosities are generally in excess with
respect to star-forming galaxies for the same FIR luminosity, the
distribution of the objects in the X-ray vs.~FIR luminosity diagram is
bounded from below from the region occupied by the star-forming
galaxies, indicating that the X-ray emission of LLAGN falling in this
border-line region could be mainly due to star formation processes,
rather than being of nuclear origin.

Our inference of using the 2--10 keV luminosity as a SFR indicator is
consistent with a recent study on Lyman-break galaxies by
\citet{nandra02} who have extrapolated the \citet{djf92} FIR/soft X-ray
relation to the hard X-ray band obtaining a SFR/2--10 keV luminosity
relation within $10\%$ of our Eq.~(\ref{pranalli-E3:eqsfrhardX2}). From
a stacking analysis of \chandra data for a sample of optically
selected Lyman-break and Balmer-break galaxies in the HDFN they find a
good agreement of the average SFR as estimated from X-ray and
extinction-corrected UV luminosities.

\citet{grimm02} have recently claimed that the luminosity function of
High Mass X-ray Binaries (HMXB) can be derived from a universal
luminosity function whose normalization is proportional to the SFR,
but only for rates $\gtrsim 1$ M$\sun$/yr, while this relation becomes
non-linear at lower rates. This appears to be somewhat in constrast
with our result that the 2--10 keV luminosity is proportional to SFR
even for rates $\lesssim 1$ M$\sun$/yr, as supported by the data for 4
galaxies in our sample. However, \citeauthor{grimm02} analysis only
refers to the contribution from HMXB; the possible contribution of
other X-ray components, which are not selected out in our data, might
explain this discrepancy.

X-ray number counts for the radio sub-mJy galaxy population are
predicted making use of the radio \lognlogs and of the X-ray/radio
correlation. The expected counts extend much below the sensitivities
of the deepest X-ray surveys (about one order of magnitude in the soft
band, one and a half for the hard band), and are within the limits set
by the fluctuations analysis in the \chandra deep fields
\citep{miyaji02a,miyaji02b}. They are also consistent with the
predictions based on the evolution of the cosmic SFR density by
\cite{ptak01}. Since the results from fluctuations analysis in the
soft X-rays suggest an excess of sources with respect to AGN synthesis
models at fluxes below $\sim 10^{-17}$ erg s$^{-1}$ cm$^{-2}$, it may
be possible that the sub-mJy galaxies represent the dominant
population in the X-rays at very faint fluxes.

The contribution to the cosmic X-ray background in the 2--10 keV band
is estimated by integration of the derived X-ray number counts.  The
contribution from galaxies detected in the deepest radio surveys is
$\lesssim 2\%$. This estimate may rise up to 11$\%$ by extrapolating
the radio counts down to 1~$\mu$Jy, or $\sim10^{-18}$ erg s$^{-1}$
cm$^{-2}$ in the X-ray band. However, since a fraction of the sub-mJy
objects may not be star-forming galaxies, these figures for the time
being should be regarded as upper limits.

The next step in understanding the physics involved in hard X-ray
emission must go through a careful analysis of {\em Chandra} and
XMM-{\em Newton} observations of both low-SFR local galaxies and
high-SFR high-redshift ones. Since the explanation of the
radio/FIR correlation is still a matter of discussion, we hope that
its extension to the X-ray band may help in clarifying this issue.

\begin{acknowledgements}
We kindly thank Joe Mazzarella for providing data before publication,
and Meri Polletta for very useful comments. We also thank an anonymous
referee whose comments have contributed to improve the presentation of
this paper. This research has made use of the VizieR database
\citep{vizier} hosted at the Centre de Donnees astronomiques de
Strasbourg (CDS), and of data obtained from the High Energy
Astrophysics Science Archive Research Center (HEASARC), provided by
NASA's Goddard Space Flight Center.  This research has been partially
supported by ASI contracts I/R/113/01 and I/R/073/01, and by the MURST
grants Cofin-00--02--36 and Cofin--01--02--8773.
\end{acknowledgements}

\appendix
\section{Notes on individual galaxies}
\subsection*{IC\,342: variability of ULXs}
In the ASCA observations, the X-ray emission of \object{IC\,342} (a
face-on spiral galaxy at 3.9 Mpc) is powered by three main sources;
two of them (source 1 and 2 according to \citealt{fabtrinch87}) are
ultraluminous X-ray binaries (ULX) while source 3 is associated with
the galactic centre. Observations with higher angular resolution
(ROSAT HRI, \citealt{bregman93}) showed that sources 1 and 2 are
point-like while source 3 is resolved in at least three sources.  Our
main concern in determining the flux of this galaxy is the variability
of the two ULX, which was assessed by a series of observations
spanning several years: IC342 was first observed by {\em Einstein} in
1980 \citep{fabtrinch87}, then by ROSAT in 1991 \citep{bregman93},
by ASCA in 1993 \citep{okada98} and 2000 \citep{kubota01}, and by \XMM in
2001.

In Table \ref{ic342ulx} we report soft X-ray fluxes for sources 1 and
2. We have chosen the soft X-ray band due to the limited energy band
of both {\em Einstein} and ROSAT; the fluxes observed with these
satellites were obtained from the count rates reported in
\citet{fabtrinch87} and \citet{bregman93} assuming the powerlaw and
multicolor disk examined in \citet{kubota01} for source 1 and 2
respectively; we take ASCA 1993 and 2000 fluxes from \citet{kubota01}.
\XMM archival observations were reduced by us with SAS 5.3 and the
latest calibrations available.

Source 1 was in a low state ($F_{0.5-2} \sim 3-5\cdot 10^{-13}$ erg
s$^{-1}$) during the 1980, 1991, 2000 and 2001 observations, and in a
high state ($F_{0.5-2} \sim 16\cdot 10^{-13}$ erg s$^{-1}$) during the
1993 observation. The broad-band (0.5-10 keV) spectrum changed, its
best-fit model being a disk black-body in 1993 and a power-law in 2000
and 2001.  Source 2 has also shown variability, its 0.5-2.0 keV flux
oscillating between $0.52\cdot 10^{-13}$ (ASCA 2000) and $2.56\cdot
10^{-13}$ (ASCA 1993) erg s$^{-1}$; the main reason for this
variability being the variations in the strongly absorbing column
density, which was $9.9\cdot 10^{21}$ cm$^{-2}$ in 1993 and $18\cdot
10^{21}$ cm$^{-2}$ in 2000. The spectrum was always a power-law.

The high state for source 1 seems thus to be of short duration, and we
feel confident that its time-averaged flux may be approximated with
its low state flux. We thus choose to derive our flux estimate for
IC\,342 from the ASCA 2000 observation, estimating the variation for
the total flux of the galaxy caused by source 2 variability to be less
than 10\%.

% ~~~~~~~~~~~~~~~~~~~~~~~~~~~~
% TABELLA: SRC 1 & 2 IN IC 342
% ~~~~~~~~~~~~~~~~~~~~~~~~~~~~
\begin{table} \label{ic342ulx}
\centering\begin{tabular}{clccc}
& & \multicolumn{2}{c}{\sc Src 1 Flux} &{\sc Src 2 Flux} \\
{\sc Year} &{\sc Mission} &{\sc bb} &{\sc po} &{\sc po} \\
\hline
1980 &{\em Einstein}   &2.7 &4.1  &0.85   \\
1991 &ROSAT            &3.3 &2.7  &1.3    \\
1993 &ASCA             &16  &     &2.6    \\
2000 &ASCA             &    &5.0  &0.52   \\
2001 &XMM-{\em Newton} &    &4.1  &1.3    \\
\hline
\end{tabular}
\caption{0.5--2.0 keV fluxes (in $10^{-13}$ erg s$^{-1}$ cm$^{-2}$) for sources 1
and 2 in IC\,342 for a blackbody (bb) and a power-law (po) model.}
\end{table}

\subsection*{Variability in M82}
{%\bf
Hard (2--10 keV) X-ray variability in M82 was reported in two
monitoring campaigns with ASCA (in 1996, \citealt{ptak99}) and RXTE
(in 1997, \citealt{rephaeli02}). M82 was found in ``high state''
(i.e.~$4\cdot 10^{-11} \lesssim f_{2-10} \lesssim 7\cdot 10^{-11}$) in
three out of nine observations with ASCA and in 4 out of 31
observations with RXTE. In all the other observations it was in a ``low
state'' ($2\cdot 10^{-11}\lesssim f_{2-10} \lesssim 3.5\cdot
10^{-11}$). A low flux level was also measured during the observations with other
experiments: HEAO~1 in 1978 \citep{griffiths79};
{\em Einstein} MPC in 1979 \citep{watson84};
EXOSAT in 1983 and 1984; BBXRT
in 1990; ASCA in 1993 \citep{tsuru97}; BeppoSAX in 1997
\citep{cappi99}; \chandra in 1999 and 2000, \XMM in 2001.  No
variability was instead detected in the 0.5--2.0 keV band
\citep{ptak99}.

The high state has been
of short duration: less than 50 days in 1996, when it was observed by
ASCA, %(M82 was in low state on 23.3.1996, then in high state on 15.4,21.4,24.4,
%and in low state on 13.5)
 and less than four months in 1997.
%(RXTE detected M82 in low state on 21.7.1997, in high state on 13.8,24.9,2.11,25.11; BeppoSAX
%detected M82 again in low state on 6.12).
A monitoring campaign was also undertaken with {\em Chandra}, which
observed M82 four times between September 1999 and May 2000.
We reduced the archival data, and found that
the galaxy was always in a low state, with its flux slowly increasing
from $1.2\cdot 10^{-11}$ to $2.9\cdot 10^{-11}$.
%dire qualcosa sul fatto che comunque chandra non ha potuto determinare
%chi è a fare questa variabilità, anche se il candidato più probabile
%è comunque la sorgente più luminosa (quella di kaaret, e prima ancora
%di watson, e di collura)?   o va bene così?

We do not attempt a detailed analysis of the variability (see
\citealt{rephaeli02}); however, we feel confident that, given the
short duration of the high states and the fact that the difference
between high- and low state flux is about a factor 2, the
time-averaged flux of M82 can be approximated with its low state
flux. We thus choose to derive our flux estimate for M82 from the
BeppoSAX 1997 observation ($f_{2-10}=2.9\cdot 10^{-11}$ erg s$^{-1}$
cm$^{-2}$), estimating the uncertainty caused by
variability to be around $30\%$.
}

\subsection*{NGC\,3256: a case for intrinsic absorption}

\begin{figure}[]    % 3256
  \begin{center} 
       \includegraphics[width=0.49\textwidth]{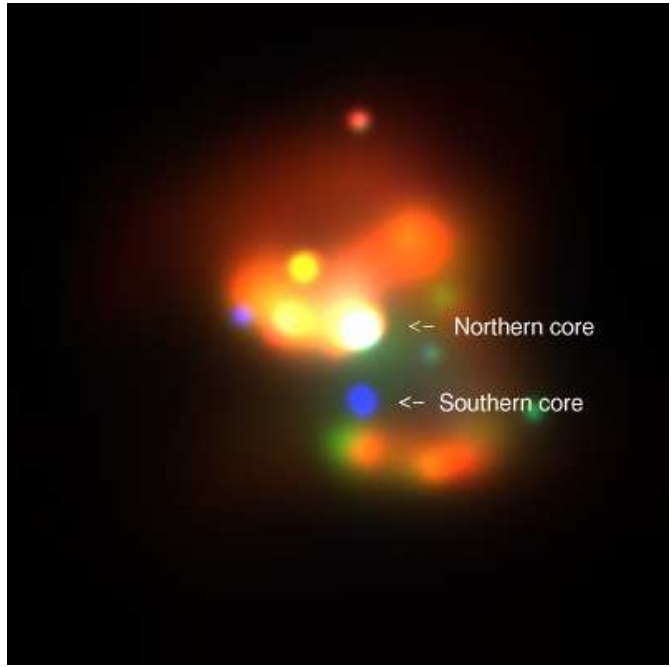}
%     \framebox[.5\textwidth]{ figura }
  \end{center}
  \caption { True color, smoothed \chandra image of the centre of NGC 3256
  (the actual colour is the sum of red, green and blue, the intensity
  of each one representing the flux in the 0.3--1.0 keV, 1.0--2.0 keV
  and 2.0--8 keV respectively); north is up, east
  is left; the distance between the two cores is $\sim 5\arcsec$.
   The two cores are the brightest sources both at 1.4 GHz and in the
  2--10 keV band, but since the southern one lies behind a dust lane,
  its soft X-ray emission is completely absorbed. While the northern
  core follows both the radio/soft X-ray and radio/hard X-ray relations
  (Eq.~\ref{eq:radiox},\ref{eq:radiox2}), the southern one only follows
  the radio/hard X-ray relation.
  \label{fig_3256} }
\end{figure}

We present the test case of \object{NGC 3256}, a luminous dusty
merger remnant included in the supplementary sample.  Detailed studies
at several wavelengths (radio: \citealt{norrisforbes95}, IR:
\citealt{kotilainen96}, optical: \citealt{lipari}, X-ray:
\citealt{moran99}, \citealt{lira}) have shown that the energetic
output of this galaxy is powered by star formation occurring at
several locations, but mainly in the two radio cores discovered by
\citet{norrisforbes95} and also detected with {\em Chandra}
(Fig.~\ref{fig_3256}). 

The 3 and 6\,cm radio maps \citep{norrisforbes95} reveal two
distinct, resolved (FWHM $\sim1.2\arcsec$) nuclei and some fainter
diffuse radio emission. Separated by $5\arcsec$ in declination, the
two cores dominate the radio emission, the northern one being
slightly (15\%) brighter. They share the same spectral
index ($\alpha\sim 0.8$) and have similar 2--10 keV fluxes.  Both of them
follow the radio/hard X-ray correlation, while only the northern one
follows the radio/soft X-ray correlation.  At other wavelengths the
northern core is the brightest source in NGC 3256, while the southern
one lies behind a dust lane and is only detected in the far infrared
($\lambda > 10\mu$m), as clearly shown in the sequence of infrared
images at increasing wavelengths in \citet{kotilainen96}.

Although the southern core appears as a bright source in the hard
X-rays ($E>2$ keV), there are not enough counts to allow an accurate
spectral fitting. However, it is still possible to constrain the
absorbing column density by assuming a template spectrum, such as a
simple power-law or the spectrum of the northern core, leading (after
standard processing of the {\em Chandra} archival observation of NGC
3256) to an intrinsic $N_{\rm H} = (2.2\pm 1.0) \cdot 10^{22}$
cm$^{-2}$ ($N_{\rm H, Gal}=9.5\cdot 10^{20}$ cm$^{-2}$), fully
consistent with the $A_{\rm V}=10.7$ estimated by \citet{kotilainen96}
from infrared observations. We note that 
84\% of the 2--10 keV flux and only 10\% of the 0.5--2 keV one
are transmitted through this column density. Thus, while
the flux loss in the hard band is still within the correlation
scatter, the larger loss in the soft band throws the southern nucleus
off the correlation of Eq.~(\ref{eq:radiox}).

%\footnote{Assuming $A_{\rm V}=4.5\cdot 10^{22} N_{\rm H}$.}

\end{document}